# Trajectory and Policy Aware Sender Anonymity in Location Based Services

Alin Deutsch[1] Richard Hull[2] Avinash Vyas[3] Kevin Keliang Zhao[1]

[1]University of California, San Diego, {deutsch, kezhao}@cs.ucsd.edu
[2]IBM T.J. Watson Research Center, hull@us.ibm.com
[3] Bell Labs Research, Alcatel-Lucent, vyas@research.bell-labs.com

## ABSTRACT

We consider Location-based Service (LBS) settings, where a LBS provider logs the requests sent by mobile device users over a period of time and later wants to publish/share these logs. Log sharing can be extremely valuable for advertising, data mining research and network management, but it poses a serious threat to the privacy of LBS users. Sender anonymity solutions prevent a malicious attacker from inferring the interests of LBS users by associating them with their service requests after gaining access to the anonymized logs. With the fast-increasing adoption of smartphones and the concern that historic user trajectories are becoming more accessible, it becomes necessary for any sender anonymity solution to protect against attackers that are trajectory-aware (i.e. have access to historic user trajectories) as well as policy-aware (i.e they know the log anonymization policy). We call such attackers TP-aware.

This paper introduces a first privacy guarantee against TP-aware attackers, called *TP-aware sender k-anonymity*. It turns out that there are many possible TP-aware anonymizations for the same LBS log, each with a different utility to the consumer of the anonymized log. The problem of finding the optimal TP-aware anonymization is investigated. We show that trajectory-awareness renders the problem computationally harder than the trajectory-unaware variants found in the literature (NP-complete in the size of the log, versus PTIME). We describe a PTIME l-approximation algorithm for trajectories of length $l$ and empirically show that it scales to large LBS logs (up to 2 million users).

## 1. INTRODUCTION

A Location-based service (LBS)[7] is an information or entertainment service, accessible with mobile devices through the mobile network and utilizing the geographic location of the mobile device (e.g. "find the nearest gas station", "Thai restaurant", "hospital"). Recently, the availability and usage of Location-based service has increased significantly because the location of mobile devices can be computed automatically (without any input from the user) by the wireless network (via triangulation of mobile device signal) or by the mobile devices themselves (via the embedded GPS chip).

Most of the popular Location-based services such as *Facebook Places* [2], *FourSquare* [3], *Gowalla* [5] and *Loopt* [6] log the LBS requests sent by their users. The data retention policies of these LBSs have provisions that describe this intent. An LBS request log is of great value to advertisers and researchers as it can be used to answer queries such as "find the requests sent by users that move from location *A* to location *B*" or "requests sent by the same user over a period of time". But an LBS request log may also contain sensitive requests that the user wishes to keep private (e.g. for the local campaign headquarter of a given political party, spiritual center for a given religion, etc.). In the event that an attacker gains access to the LBS request log, the sender's privacy is at risk.

In this paper we investigate how to anonymize the LBS request log so as to protect the identity of the LBS request senders even if the log falls in the hands of attackers who also gain access to *a)* the sequence of location-timestamp pairs (a.k.a. *trajectory*) visited by the mobile users for the duration the LBS requests are logged and *b)* the anonymization policy used to provide this protection. Assumption *b)* is based on a well-accepted principle of designing a private and secure system - "The design is not a secret" [27]. Assumption *a)* is a realization of the fact that an attacker can obtain the locations visited by the users from many sources, including the wireless service provider, or location computing servers such as SkyHook [8], or user surveillance. A recent article in the *Wall Street Journal* [9] and a joint study [17] by *Intel Labs*, *Penn State*, and *Duke* University provide evidence that advertisers are logging the trajectories of the mobile device users. The attacker may gain access to this information via hacking, financial agreement and subpoena.

In the LBS context, the best-studied identity protection measure is known as *sender k-anonymity* [20, 19, 24, 20, 16], which is intended to guarantee that the content of an LBS request and the precise location of the users are insufficient to distinguish among the actual sender and k-1 other possible senders. This guarantee is targeted towards the LBS request sent by the user at a given instant of time. The underlying model does not consider the *sequence* of LBS requests sent by the user over time. We refer to this privacy guarantee as *snapshot sender k-anonymity*, and any solution enforcing it as *snapshot k-anonymization*. As shown below, snapshot sender k-anonymity protects only against attackers who are unaware of the user trajectories, i.e. treating requests at different instants as independent even if they actually originate from the same user. Typical snapshot anonymization algorithms [19, 24, 20, 16] are based on hiding the sender's precise location $l$ in the request, substituting instead a $cloak$, i.e. a region containing $l$. The cloak is usually chosen from among regions of a pre-defined shape (circular, rectangular etc.), to include locations of at least k-1 other mobile users. We refer to the cloak selection policy as *snapshot k-anonymous policy*. We illustrate a snapshot 2-anonymous policy next.

EXAMPLE 1. *Figure 1 and Figure 2 show the location of five*

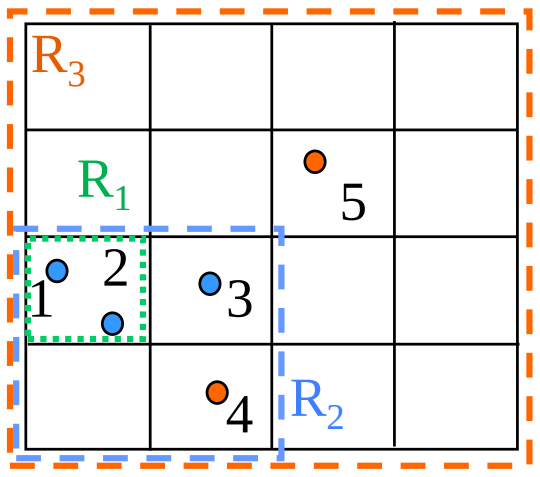

Figure 1: User locations at $t_1$

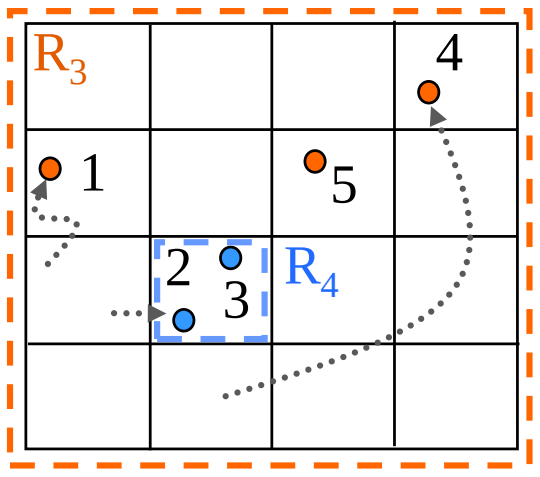

Figure 2: User locations at $t_2$

| | |
|---|---|
| $P_1$ | $1 \to R_2, 2 \to R_2, 3 \to R_2, 4 \to R_3, 5 \to R_3, \cdots$ |
| $P_2$ | $1 \to R_3, 2 \to R_4, 3 \to R_4, 4 \to R_3, 5 \to R_3, \cdots$ |

Table 1: Snapshot policy-aware 2-anonymous policies

*users at two instants $t_1$ and $t_2$. Table 1 shows the cloak selection policies $P_1$ and $P_2$ that select cloaks from the quadrants of a static quad-tree based partitioning of the geographic space. Suppose at instant $t_1$, user 1 sends an LBS request $L_1$. Policy $P_1$ anonymizes $L_1$ by substituting the location in the request with the cloak $R_2$ (shown in Figure 1). Note that $R_2$ includes the location of users 1, 2, 3 and a request sent by any one them is anonymized by $P_1$ using the same cloak $R_2$. Thus when an attacker, who has access to the user locations at $t_1$ and policy $P_1$, observes the anonymized request with cloak $R_2$, he cannot distinguish whether the sender is user 1 or 2 or 3. Thus $P_1$ provides snapshot sender 2-anonymity.*

*Suppose user 1 sends another LBS request $L_2$ at instant $t_2$. Policy $P_2$ anonymizes $L_2$ by substituting the locations in the request with cloak $R_3$ (shown in Figure 2). An attacker who has access to user locations at $t_2$ and policy $P_2$ cannot distinguish the sender among users 1, 4 and 5 since requests from all these users are anonymized using $R_3$. Thus $P_2$ provides snapshot sender 2-anonymity. Note that policy $P_2$ does not take into account the user locations at instant $t_1$ and their anonymizations using policy $P_1$ (and vice versa).* □

A natural first candidate solution to anonymizing LBS request logs is to leverage previous work on snapshot k-anonymization [19, 24, 20, 16], anonymizing for each time instant $t$ the snapshot of requests at $t$ (independently of how snapshots at other instants are anonymized). Unfortunately snapshot-by-snapshot anonymization of a request log does not provide sender k-anonymity against an attacker who has access to the user trajectories for the period the requests are logged. The next example illustrates this point.

EXAMPLE 2. *Recall the setting in Example 1 and assume that the LBS logs the user requests. To anonymize the log, the LBS uses policies $P_1$ and $P_2$. Moreover, user ids are replaced with meaningless identifiers, however in order to preserve the linkage between requests sent from the same trajectory, the same identifier is used for all requests sent by the same user.*

*Assume the anonymized request log is observed by an attacker who knows $P_1$ and $P_2$ and the user locations at $t_1$ and $t_2$. As in Example 1, the attacker can use the knowledge of the policies to limit the first request's sender to one of $\{1, 2, 3\}$ and the second request's sender to one of $\{1, 4, 5\}$. Next he uses the additional knowledge that both requests where sent from the same user trajectory: he intersects the two sets of potential senders and concludes that user 1 must be the sender, breaching sender 2-anonymity!* □

In Example 2 the attacker is able to breach sender 2-anonymity because the request log enables him to associate the two requests to the same trajectory and the snapshot 2-anonymization policy $P_2$ does not take into account the anonymization of request from user 1 at instant $t_1$ using policy $P_1$. Thus the above breach could have been avoided if either the two requests were not linked with the same trajectory in the anonymized log, or if instead of policy $P_2$ we used a policy $P_2'$ that anonymizes requests from users 1, 2 and 3 using the region $R_3$. While we are free to change the anonymization policy to preserve sender k-anonymity, we do not wish to entirely remove the association of requests with a trajectory in the anonymized LBS request log since this is valuable information for analytics. This poses an interesting challenge for the LBS: how can it publish an anonymized LBS request log that includes some form of linkage information between requests and trajectories, without jeopardizing k-anonymity of the users?

The LBS needs to ensure anonymity against an attacker who knows the user trajectories for the period the LBS requests are logged (we call the attacker *T-aware*) and who knows the "design" i.e. the policy used to pick cloaks for anonymizing requests (*P-aware* attacker). We call this problem the *offline TP*-aware sender k-anonymity problem since an LBS request $r$ can be anonymized taking into account all request in the log, including those sent after $r$. We contrast this with the problem of *online TP*-aware sender k-anonymity, in which (i) the LBS is not trusted, therefore an LBS request is anonymized before it is sent to the LBS, and (ii) the anonymization of an LBS request takes into account only the history of requests so far and cannot be altered after observing subsequent requests by the same user. We leave the problem of online *TP*-aware sender k-anonymity for future work.

In this paper, we propose a solution to the offline TP-aware sender k-anonymity problem. It consists of publishing a sequence of cloaks to anonymize the sequence of LBS requests sent by a user over a period of time. With each cloak in the sequence we associate a set of LBS requests devoid of any location and sender identity information. The LBS requests associated with a sequence $S$ of cloaks represents LBS requests by users whose trajectories pass through $S$. We thus preserve some association between LBS requests and the trajectories they were sent along (though we introduce some uncertainty as requests are not tied to a single trajectory, but rather to a "bundle" of trajectories compatible with $S$). To provide TP-aware sender k-anonymity, we choose $S$ such that at least k distinct user trajectories are anonymized to $S$.

The technical challenge we need to solve is to find, among the (exponentially) many possible ways of bundling user trajectories together, the one that results in the maximum utility for the consumer of the anonymized log. Intuitively, we are looking to minimize the cloak sizes, so as to improve the precision of the anonymized information.

**Our contributions** include the following:

**[1]** We identify and formulate the problem of offline sender k-anonymization of LBS request logs, which protects against the class of trajectory- and policy-aware attackers. We define a novel privacy guarantee, *TP-aware sender k-anonymity*.

**[2]** We study the problem of finding, among all the offline policies that provide *TP*-aware sender k-anonymity, one with the optimum utility for the consumer of the anonymized LBS log. We show that finding the optimum offline policy that uses cloaks chosen among the quadrants of a quad-tree based partition of the map is *NP-Complete*. This is significant, showing that guarding against T-aware attackers is computationally harder than against T-unaware attackers: it was shown in [16] that for such cloak types optimum snapshot k-anonymization is in PTIME.

**[3]** We show that optimum *TP*-aware sender k-anonymity is PTIME-approximable (i.e. one can always find, in polynomial time, an anonymization whose utility is within well defined bounds rela-

tive to the optimum utility). In particular, we describe a novel $l$-approximation algorithm to anonymize an LBS request log spanning user trajectories of length $l$.

**[4]** We implement and experimentally evaluate our anonymization algorithm and show that it is practical and scales well with the number of user trajectories: it takes less than 4 minutes to anonymize 2 million trajectories of length 30 for users moving around the San Francisco Bay area. This is a performant running time for an offline algorithm, especially since we show that alternate solutions are impractically slow and/or provide worse approximations.

**Paper outline**. The remainder of the paper is organized as follows. In Section 2, we describe a prevailing model of an LBS. We define offline TP-aware sender k-anonymity in Section 3 and describe our solution that uses a sequence of cloaks to preserve TP-aware sender k-anonymity while publishing some linkage information between anonymized requests and bundle of user trajectories. In Section 4 we show that finding the optimum offline policy that provides TP-aware k-anonymity is NP-hard and hence in Section 4.1 we propose a polynomial time approximation algorithm. In Section 4.1.3 we describe optimizations and our implementation of the optimized approximation algorithm. We report on the experimental evaluation in Section 5, discuss related work in Section 6 and conclude in Section 7.

## 2. LOCATION BASED SERVICES

This section introduces a prevailing model of location-based services, based on automatic computation of the location of user mobile devices. It describes various entities in the LBS ecosystem, the data flow among these entities and the data logged by them.

As shown in Figure 3 there are four core elements in the delivery of a location-based service: the user making a request, typically called the *sender*, the (wireless) Communication Service Provider, denoted as *CSP*, the location server that computes the location of the mobile device, denoted as *LS*, and the Location Based Service (LBS) provider, denoted as *LBS*. We view the CSP, Location Server and LBS provider to be trusted agents and assume that the communication between them is secure.

To access an LBS, a sender uses an application on the mobile device (typically provided by the LBS). The application fetches the location from the run-time environment on the mobile device, which in turn gets it from the location server. The location server is a specialized network component in CSP's network, known as *Mobile Positioning Center* (MPC) in the CDMA standard, that provides access to device locations for E911 [1] and other location-based services. The location can also be obtained from a service that operates outside the CSPs network and can compute the location of a mobile device using the signal strength of nearby cell-towers and WiFi access points observed on the mobile device (e.g. SkyHook [8] and Google Location Service [4]). The application then sends a service request containing the location and the specifics of information/operation requested by the sender (e.g. "car dealership in 5 mile radius from my location" or "notify me when a friend is within 1 mile from my location"). The LBS provider responds to the LBS request using the location sent with the request.

Henceforth we abstract from these details and focus on the treatment of location data and the LBS service requests in the LBS ecosystem. For simplicity of presentation (and without loss of generality), we model a geographic area as a 2-dimensional space and user's location as integer coordinates within this space.

As a mobile user moves and sends LBS requests from different locations at different times, the LBS provider logs these requests.

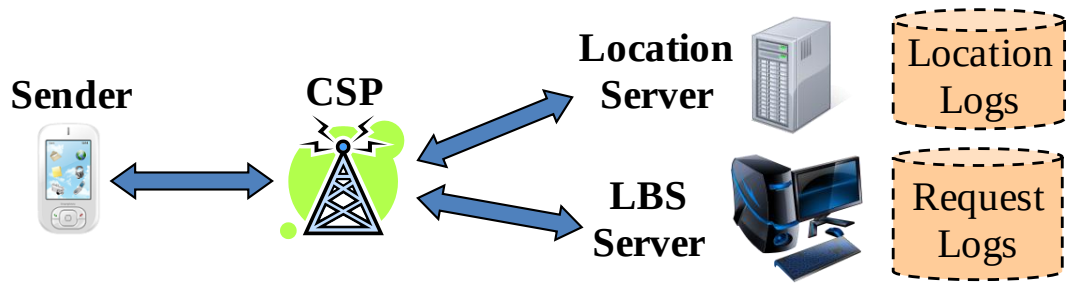

Figure 3: LBS Model

Each logged request is associated with the identifier of the device that sent the request (e.g. IP-address or MAC-address). This allows the LBS provider to identify the requests sent by the same user in the LBS request log and assemble a history of LBS requests for each user. We represent a *user history* of length $l$ as a triple

$$(uid, \langle loc_1, loc_2, \dots loc_l\rangle, \langle V_1, V_2, \dots V_l\rangle)$$

where $uid$ is the userid of the user, $loc_i$ is the location of the user at instant $i$ and $V_i$ the request sent by the user at instant $i$. (For presentation simplicity and without loss of generality, we abstract away the actual timestamps, representing them as natural numbers.) Each *request* is a set of name-value pairs containing the categories and specifics of the sought services (e.g. $\{(poi, rest)$,$(cat, ital)$, $(dist, 2mi)\}$ represents "find Italian restaurants within 2 miles of my location"). An *LBS request log* consists of a set of user histories. The *snapshot* at instant $i$ contains for each user her location at instant $i$ and the request (if any) it made at $i$.

EXAMPLE 3. *Consider an LBS request log containing the histories $u_a, u_b, u_c, u_s, u_t$ of users Alice, Bob, Carrol, Sam and Tom, whose devices have IDs $id_a$, $id_b$,$id_c$,$id_s$ and $id_t$, respectively:*

$u_a = (id_a, \langle(1,2),(1,3)\rangle, \langle v_1, v_2\rangle)$
$u_b = (id_b, \langle(1,2),(2,2)\rangle, \langle v_3, _\rangle)$
$u_c = (id_c, \langle(2,2),(2,2)\rangle, \langle _, v_4\rangle)$
$u_s = (id_s, \langle(2,1),(4,4)\rangle, \langle v_5, _\rangle)$
$u_t = (id_t, \langle(3,3),(3,3)\rangle, \langle _, v_6\rangle)$

*From this log we can read that at instant 1, the LBS logs the requests $v_1$, $v_3$ and $v_5$ sent by the users Alice, Bob and Sam (respectively). Alice was at that time at location $(1,2)$, given as coordinates in the 2-dimensional map. At instant 2, the LBS logs the requests $v_2$, $v_4$ and $v_6$ sent by Alice, Caroll and Tom (respectively). Alice had by now moved to location $(1,3)$. _ stands for "no request" (Bob made no request at instant 2).* □

The set of user history objects contains very useful data for researchers and advertisers since it can be used to answer queries such as "the requests sent by users that move from location *A* to location *B*". As mentioned in Section 1, previous work on sender anonymity in an LBS setting has focused on anonymizing snapshots independently of each other [20, 19, 24, 20, 16]. Example 2 showed that leveraging such work to user histories by anonymizing them snapshot-by-snapshot leads to privacy breach. A more holistic anonymization is called for.

## 3. A NOVEL PRIVACY GUARANTEE

In this section we describe our approach for anonymizing the set of user histories to provide sender k-anonymity against the class of attackers who are aware of user trajectories and the anonymization algorithm, in the sense that the attacker cannot reduce the set of possible suspects to less than $k$. The anonymization preserves linkage information between LBS requests and trajectories to an extent that does not pose any risk to the sender k-anonymity of the users.

**Bundle** The key concept we introduce to model an anonymized user history is called a "bundle". Intuitively, a bundle object corresponds to a set of user histories bundled together. It lists a sequence of cloaks such that the cloak at instant $i$ contains the locations at instant $i$ of all users in the bundle. The bundle also lists for each instant $i$ the set of requests issued at instant $i$ by the bundled users.

More formally, a *bundle* is a tuple $(bid, \langle r_1, \ldots, r_l\rangle, \langle s_1, \ldots s_l\rangle)$, where $bid$ is a unique bundle identifier, $r_1, \ldots, r_l$ is a sequence of $l$ cloaks, and $s_1, \ldots, s_l$ is a sequence of $l$ sets of LBS requests. A *cloak* is a 2-dimensional region (e.g. $[(x_1, y_1), (x_2, y_2)]$ for axis-parallel rectangles, where $(x_1, y_1)$ and $(x_2, y_2)$ are the coordinates of the lower-left and upper-right corners of the rectangle). Recall from Section 2 that a request is a set of name-value pairs devoid of any identifier or location information.

EXAMPLE 4. *The following are examples of bundles that use the cloaks shown in Figure 1 and Figure 2 and the requests described in Example 3.*

$b_1$ = *(1,* $\langle R_2, R_3\rangle$, $\langle\{v_1, v_3, v_5\}, \{v_2, v_4\}\rangle$*)*
$b_2$ = *(2,* $\langle R_3, R_3\rangle$, $\langle\{v_1, v_3, v_5\}, \{v_2, v_4, v_6\}\rangle$*)*
$b_3$ = *(3,* $\langle R_2, R_4\rangle$, $\langle\{v_3\}, \{v_4\}\rangle$*)*
$b_4$ = *(4,* $\langle R_3, R_4\rangle$, $\langle\{v_3\}, \{v_4\}\rangle$*)* □

DEFINITION 1. *[**Masking**] Given a user history* $u = (uid, \langle loc_1, loc_2, \ldots loc_l\rangle, \langle V_1, V_2, \ldots V_l\rangle)$ *and a bundle* $b = (bid, \langle r_1, \cdots, r_l\rangle, \langle s_1 \cdots s_l\rangle)$*, we say that* $b$ masks $u$ *if for each* $i \in [1, \ldots l]$*,* $loc_i \in r_i$ *and* $V_i \in s_i$*.*

EXAMPLE 5. *Assuming unit length for the smallest squares in Figure 1, bundle* $b_1$ *in Example 4 masks the user histories* $u_a$*,* $u_b$*,* $u_c$ *and* $u_s$ *from Example 3. Similarly bundle* $b_3$ *masks the user histories* $u_b$ *and* $u_c$*.* □

Instead of publishing the set of user histories, our proposal is that the LBS publish a set of bundles that masks the user history objects in its log. Note that within a bundle, the association of the LBS requests with the trajectory of the actual sender is obfuscated since for any two distinct instants $i, j$ and any requests $V_i, V_j$ the bundle only states that their sender locations belong to the cloaks $r_i, r_j$, but not whether they were actually sent by the same user.

Insuring sender k-anonymity consists in choosing bundle objects such for each request in the bundle, an attacker cannot limit the set of potential senders to less than $k$. We formalize this next.

**Anonymization Policy** We define an *anonymization policy* as a function $P$ that, given a set of user history objects $U$, associates to each user history object $u$ a bundle object $b$ (denoted $P(U, u) = b$) such that $b$ masks $u$. We sometimes write $P(u) = b$ when the LBS log $U$ is clear from the context.

EXAMPLE 6. *The following anonymization policy* $P_3$ *anonymizes the 5 user history objects in Example 3 using the bundle objects shown in Example 4.* $P_3(u_a) = b_1$*,* $P_3(u_b) = b_3$*,* $P_3(u_c) = b_3$*,* $P_3(u_s) = b_1$*,* $P_3(u_t) = b_2$*.* □

## 3.1 TP-aware Sender k-Anonymity

We next define our novel privacy guarantee. To do so we need to formalize the class of attackers who are aware of user trajectories and the anonymization policy.

**Attacker Model** We target a strong information-theoretic definition of privacy therefore we model the attacker as a function taking certain input to launch the attack, with no bounds on the computational resources expended during the attack. The only assumptions are on what input the function takes (intuitively, the information that the attacker sees). The input comprises:

- the anonymity degree $k$;
- the specific anonymization policy $P$ used by the LBS;
- the trajectory of all the users (the first and second components of each user history triple);
- the published bundles.

We refer to the class of such attackers as *Trajectory-aware and Policy-aware* (in short, TP-aware) attackers. The attack function models the following attack: starting from the observation of a set $B$ of bundle objects, the knowledge of trajectories of all the users and the anonymization policy $P$, the attacker reverse engineers $P$ to obtain the possible user histories that are anonymized by $P$ to bundles in $B$.

We are now ready to define TP-aware sender k-anonymity. Intuitively, we consider it a breach of sender k-anonymity if for any bundle $b$ the attacker succeeds in reducing the number of candidate user histories that can possibly be anonymized to $b$, to less than k. Therefore, our privacy guarantee ensures that for each observed bundle object $b$ there are at least $k$ user histories that are anonymized to $b$ under the chosen policy.

DEFINITION 2. *[**TP-aware Sender k-anonymity**] Let* $P$ *be an anonymization policy and* $U$ *be a set of user histories. Let* $B$ *be the set of bundles obtained using the policy* $P$*. We say that* $B$ *provides* TP-aware sender k-anonymity *for* $U$ *if for each bundle* $b \in B$ *there are at least k distinct user histories in* $U$ *that are anonymized to* $b$ *under* $P$*. We say that policy* $P$ *provides* TP-aware sender k-anonymity, *if for every set of user histories* $U$*, the set of bundles* $\{P(u)|u \in U\}$ *provides* TP*-aware sender k-anonymity to* $U$*.*

Note that even with unlimited computational resources, the best the attacker can hope to achieve is to exactly reverse engineer the inverse image of the published bundles under $P$ (as opposed to just approximating it). Since $P$ is defined such that the inverse image contains at least $k$ possible user histories for each bundle, even the exact inversion of $P$ does not breach privacy.

We next show a policy that breaks *TP*-aware sender 2-anonymity.

EXAMPLE 7. *Policy* $P_3$ *of Example 6 anonymizes the set of user histories* $\{u_a, u_b, u_c, u_s, u_t\}$ *shown in Example 3 to the set of bundles* $\{b_1, b_2, b_3\}$ *shown in Example 4. When the TP-aware attacker observes* $b_1$*, he tries to reverse engineer the user-history objects that could have anonymized to it. He finds two candidates,* $u_a$ *and* $u_s$*, corresponding to users* $Alice$ *and* $Sam$*. Similarly for* $b_3$*, there are 2 users* $u_b$ *and* $u_c$ *that could be anonymized to* $b_3$*. In contrast, when the attacker observes* $b_2$*, there is only one user history* $u_t$ *that is anonymized to* $b_2$ *under* $P_3$*. Thus* $P_3$ *does not provide* TP*-aware sender 2-anonymity.*

We next illustrate a policy that does provide *TP*-aware sender 2-anonymity.

EXAMPLE 8. *For the five user histories in Example 3, we describe the following anonymization policy* $P_4$ *that uses the bundles shown in Example 4.* $P_4(u_a) = b_2$*,* $P_4(u_b) = b_3$*,* $P_4(u_c) = b_3$*,* $P_4(u_s) = b_2$*,* $P_4(u_t) = b_2$*. There are at least 2 user histories anonymized by* $P_4$ *to each one of the published bundles,* $b_2$ *and* $b_3$*. When the attacker observes the published bundles, he tries to reverse engineer* $P_4$*, but finds at least 2 users for each of* $b_2$ *and* $b_3$*. Hence* $P_2$ *provides TP-aware sender 2-anonymity.* □

# 4. OPTIMUM-COST ANONYMIZATION

For the same set of user-history objects there may exist several anonymization policies that provide *TP*-aware sender k-anonymity, raising the obvious question of which one to use. In this section we address the problem of finding the k-anonymous policy of highest utility to the consumers of the published log. Prior work [16, 20, 19, 24] on snapshot sender k-anonymity proposes that one way to maximize utility is to minimize the area of the cloaks. For the log of LBS requests, an analogous measure would be to minimize the sum of the cloak areas used in the bundles.

**Cost of a bundle.** We introduce the cost of a bundle to quantitatively measure utility (maximum utility means minimum cost). Given a bundle $b = (bid, \langle r_1 \cdots r_l \rangle, \langle s_1 \cdots s_l \rangle)$, we define the *cost* of bundle $b$ as the sum of the areas of the cloaks in its cloak sequence: $Cost(b) = \sum_{i=1}^{l} area(r_i)$. Given a collection $U$ of user objects and an anonymizing policy $P$, we define the *cost* of $P$ for anonymizing $U$ as $Cost(P, U) = \sum_{u \in U} Cost(P(u))$.

**Optimum policy** We next focus on the problem of finding the optimum (minimum cost) policy that provides *TP*-aware sender k-anonymity to a given set of user-history objects.

Notice that the TP-aware sender k-anonymity guarantee is at least as computationally hard to enforce as its P-aware snapshot (T-unaware) version (the latter is a special case of the former for trajectory length 1). It is therefore natural to avoid settings in which snapshot anonymization is already intractable. For P-aware snapshot k-anonymity, it was shown in [16] that the complexity of finding the optimum policy depends crucially upon the type of cloaks used for anonymization. For instance, finding the optimum policy among all the policies that use circular cloaks is NP-hard (in the number of users) [16], even if the cloak centers can be chosen only from a given set of points (e.g. public landmarks such as libraries, train stations or cell towers) and the only choice is on the length of the cloak radius. In contrast, one can find an optimum snapshot policy in polynomial time if the cloaks are chosen among the quadrants of a quad tree [16]. These results suggest that, for a chance at practically feasible anonymization, our investigation would best focus on quad-tree based cloaks. This is indeed the case as shown in the next theorem.

**Anonymization using Circular cloaks.** Let $U$ be a set of user-history objects and $SC$ be a set of points in the 2-dimensional space that contains the trajectories of the users. We define *circular cloak sequence* as a sequence of cloaks where each cloak is centered at some point from $SC$, with no restriction on the radius. Let $\mathcal{P}$ be the set of all those policies that use circular cloak sequence in the bundles for anonymizing user-history objects. The problem of *Optimum Offline* TP-*aware k-anonymization with Circular cloaks* is to find a policy in $\mathcal{P}$ that minimizes the cost of anonymizing $U$.

(**Extended Version) Theorem** 4. *Optimum Offline* TP-*aware k-anonymization with Circular cloaks is NP-hard.*

The quad tree is a well-known structure for organizing spatial data, and it has been used in a number of anonymization solutions [19, 24, 16] for snapshot sender k-anonymity. More important, this is the only cloak class for which optimum P-aware snapshot k-anonymization is known to be PTIME-computable [16].

**Anonymization using Quad-cloaks** For the remainder of the paper, we consider policies that use cloaks picked from among the quadrants of a quad-tree partitioning of the geographic region. The root node of the quad-tree represents the entire region (assumed square shaped, without loss of generality) which is then partitioned into 4 equal square quadrants, each of whom represent a child node of the root. Each quadrant is then again divided into 4 equal sub-quadrants that correspond to grandchildren of the root. This four-way splitting goes on recursively until the desired level of granularity for the minimum region is reached. Figure 4 shows a part of a quad-tree based partitioning: region $R_8$ represents a quadrant in the quad-tree that is divided into 4 equal sub-quadrants (e.g. $R_4$). The sub-quadrant $R_4$ is further divided into $R_0$, $R_1$, $R_2$, and $R_3$.

Given a quad-tree representation $Q$ of a region, we refer to a sequence of cloaks, where each cloak is one of the quadrants of $Q$, as a *quad-cloak sequence*. For instance, $\langle R_0, R_3 \rangle$ is a quad-cloak sequence of length 2 that uses the quadrants of the quad-tree in Figure 4. A policy that anonymizes user histories using bundles with quad-cloak sequence is referred to as a *quad-cloak policy* (since we consider only such policies for the remainder of the paper, we will drop the qualifier whenever convenient).

**Optimum quad-cloak policy.** Given a quad-tree $Q$ and a set $U$ of user histories, there exist several quad-cloak policies that can be used to anonymize $U$. We show that their number is exponential. Assume that $U$ comprises $n$ user histories, each of length $l$, and the quad-tree $Q$ is of height $h$. For any location in the trajectory of a user history, there are $h$ cloaks in $Q$ that mask it (all the cloaks from leaf to root in $Q$). Therefore, for a trajectory of length $l$, there are $h^l$ different quad-cloak sequences masking it. There are hence $h^{(nl)}$ different ways of anonymizing the $n$ histories in $U$ using quad-cloak sequences from $Q$ (although not all of them provide *TP*-aware sender k-anonymity).

The problem of *optimum offline* TP-*aware sender k-anonymity with quad-cloaks* is to find, given LBS log $U$, a quad-cloak bundle $B$ that has the minimum cost of anonymizing $U$. Clearly a brute-force search among all $h^{(nl)}$ quad-tree policies would take exponential time in the log size. As shown by our next result one cannot hope for PTIME (unless P = NP).

THEOREM 1. *Optimum offline* TP-*aware sender k-anonymity with quad-cloaks is NP-complete (in the size of the LBS log).*

The significance of this result is that it shows that providing optimum TP-aware sender k-anonymity is strictly a harder problem than the optimum snapshot k-anonymity studied in prior work.

## 4.1 Approximation Algorithm

The next best thing in lieu of a polynomial-time optimum solution is to find a polynomial time approximation algorithm with bounded approximation factor. We show such an algorithm next.

At high level, we proceed as follows. We restrict the choices of cloak sequences that a policy can use, to a subset of all the possible choices of quad-cloak sequences. This amounts to identifying a subset $S'$ of the set $S$ of all possible quad-cloak policies. The subset $S'$ is chosen such that an optimum quad-cloak policy relative to $S'$ can be found in polynomial time, and that this policy's cost is within a bounded factor of the optimum quad-cloak policy in $S$.

Our algorithm utilizes a structural relationship that exists between quad-cloak sequences of a given length $l$.

**1-step Generalization and Generalization Graph** Let $Q$ be a quad-tree and $s$ be a quad-cloak sequence of length $l$ that uses quadrants of $Q$. Let $s'$ be a quad-cloak sequence obtained by replacing one of the cloaks in $s$ with its parent in $Q$. We refer to $s'$ as *1-step generalization* of $s$. The 1-step generalization relation induces a directed acyclic graph over all the quad-cloak sequences of a given length $l$ obtained using a quad-tree $Q$. We refer to this graph as the *Generalization Graph* (G-graph for short). Figure 6 shows part of the G-graph induced by 1-step generalization on the quad-cloak sequences of length 2 that use quadrants of the quad-tree shown in Figure 4.

In a G-graph of length $l$ it is easy to observe that a trajectory of length $l$ masked by a quad-cloak sequence $s$ is also masked by the 1-step generalization of $s$. We refer to this property as the *containment* property. As an example, consider the trajectory of the user $a$ shown in Figure 5. This trajectory is masked not only by the quad-cloak sequence $\langle R_0, R_3 \rangle$ shown in Figure 6, but also by the sequences corresponding to its ancestors in the G-graph (e.g. $\langle R_0, R_4 \rangle$ and $\langle R_4, R_4 \rangle$).

Our approach to finding a subset of quad-cloak policies reduces to finding a tree-shaped subgraph of the G-graph: Given a G-graph $G$ of length $l$ and an LBS log $U$, let $\mathcal{P}_G$ be the set of all the policies that use quad-cloak sequences from $G$. The problem of optimum offline $TP$-aware sender k-anonymity with quad-cloaks is to find the optimum policy in $P_G$. Since this is $NP$-hard, we identify a subspace $T$ of the $G$, and find the optimum policy in the set $\mathcal{P}_T$ of all policies that use cloak sequences from $T$. The choice of $T$ is such that the optimum policy can be found in PTIME and its cost is a bounded approximation of the optimum policy in $P_G$.

**G-tree of a G-graph** Given a G-graph $G$, we define a *Generalization tree (G-tree)* of $G$ as a tree $T$ in which every node has bounded degree and that preserves the ancestor-descendant relationship of $G$. Formally: a) The nodes of $T$ are a subset of the nodes in $G$. b) If $y$ is the parent of $x$ in $T$, then $y$ must be a ancestor of $x$ in $G$. c) Each node in $T$ has a finite bounded degree $f$ (i.e. each non-leaf node has at most $f$ children).

Conditions a) and b) ensure that the set $\mathcal{P}_T$ of all the policies that use the cloak sequences in $T$ is a subset of all the quad-cloak policies $\mathcal{P}_G$. In addition, property b) above also preserves the G-graph containment in the corresponding G-tree. As a result any trajectory masked by a node in $T$ is also masked by its parent in $T$. As described next, this property along with condition c) is key in finding a polynomial time approximation solution.

Note that using the above definition, one can obtain multiple G-trees corresponding to a G-graph. The choice of a G-tree dictates the bounded approximation factor and the complexity of the algorithm that achieves the bound. We address the issue of identifying a G-tree with bounded approximation factor in Section 4.1.2. We first describe in Section 4.1.1 a generic algorithm that takes as input a G-tree $T$ and finds in PTIME the optimum policy w.r.t. to $\mathcal{P}_T$.

### 4.1.1 Min-Cost Policy in Any G-tree

The algorithm exploits two unique properties of the policies in $\mathcal{P}_T$. Using the first property we define equivalence classes of policies such that all equivalent policies have the same cost and anonymize the same number of trajectories to each quad-cloak sequence in $T$. This allows us, instead of exploring a search space of policies, to explore a smaller search space of policy equivalence classes. Even though there are fewer equivalence classes than policies, the total number of choices is still exponential (in the number of cloak sequences in $T$). The second property allows us to use a divide and conquer strategy to prune and search in polynomial time the exponential search space of equivalence classes and find the one corresponding to the optimum policy.

**Property 1: Cost of a policy is determined by the number of trajectories anonymized to each node in $T$.** For a policy in $\mathcal{P}_T$, the property of being *TP*-aware sender k-anonymous and the cost of the anonymization depends upon *how many* trajectories are anonymized by each node $n$ in $T$, being indifferent to *which* particular trajectories are anonymized to $n$.

EXAMPLE 9. *Consider trajectories a and b shown in Figure 5. Let $P_1$ and $P_2$ be two anonymization policies that anonymize trajectories a and b as shown in Figure 7 and Figure 8. $P_1$ anonymizes a to cloak sequence $\langle R_0, R_4 \rangle$ and b to $\langle R_4, R_4 \rangle$, whereas $P_2$ anonymizes a to $\langle R_4, R_4 \rangle$ and b to $\langle R_0, R_4 \rangle$. Except for this difference, all the other trajectories are anonymized identically in $P_1$ and $P_2$. Since $Cost(P_1(a)) = Cost(P_2(b))$ and $Cost(P_1(b)) = Cost(P_2(a))$, we have $Cost(P_1(a)) + Cost(P_1(b)) = Cost(P_2(b)) + Cost(P_2(a))$ and the costs of $P_1$ and $P_2$ are identical.* □

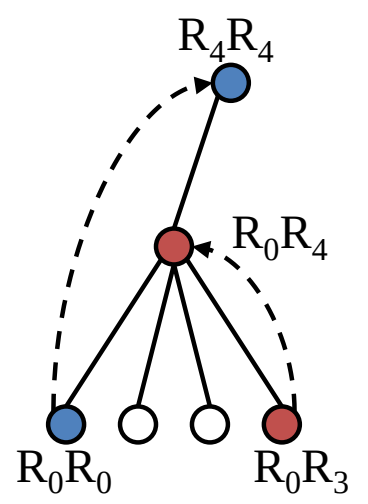

Figure 7: Policy $P_1$

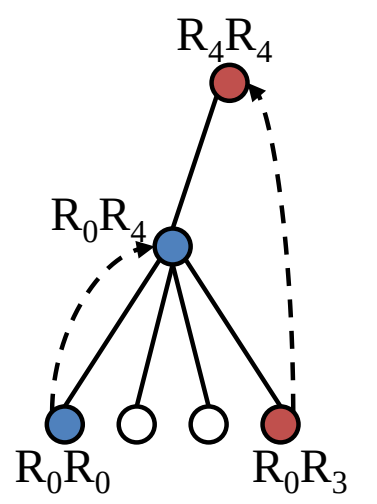

Figure 8: Policy $P_2$

We formalize this observation as an equivalence relation among policies in $\mathcal{P}_T$ that use quad-cloak sequences in G-tree $T$. Two policies in $\mathcal{P}_T$ are *equivalent* for a given set of trajectories if every node in G-tree $T$ anonymizes the same number of trajectories under both policies.

LEMMA 1. *If policies $P_1, P_2$ are equivalent for a G-tree $T$, then (a) $P_1$ and $P_2$ have the same cost; and (b) $P_1$ provides* TP-*aware sender k-anonymity on $T$ if and only if so does $P_2$.*

We exploit equivalence to replace the search space of policies in $\mathcal{P}_T$ with the smaller space of equivalence classes. We represent equivalence classes using a *Configuration* function.

**Configuration** The function *Configuration* is defined to keep track of the number of trajectories anonymized by each node $m$ in a G-tree $T$. For technical convenience, this is done by equivalently tracking for each node $m$ the number of trajectories that are masked by $m$ yet are *not* anonymized using $m$ or any of its descendants. We refer to these trajectories as *passed up* (the responsibility of anonymizing them is passed up to $m$'s ancestors).

DEFINITION 3. *[**Configuration**] Let $U$ be a set of trajectories and $T$ be a G-tree rooted at $r$. Let $d(m)$ denote the total number of user trajectories that are masked by the cloak sequence represented by node $m$. A* Configuration *$C$ is a function from nodes of $T$ to natural numbers, such that (i) for every leaf node $m$, $C(m) \leq d(m)$; and (ii) for every internal node $q$, $C(m) \leq \sum_{i=1}^{f} C(m_i)$, where $m$ has $f$ children $m_1, \ldots, m_f$. We say that $C$ is* complete *if $C(r) = 0$.*

Condition *(i)* in the above Definition 3 restricts a configuration to represent only masking policies and *(ii)* represents the fact that a trajectory can be anonymized to only one cloak sequence. Note that by Lemma 1(a), all policies in the equivalence class represented by a configuration $C$ have the same cost. We call this the cost $Cost_c$ of the configuration $C$. We can compute this cost directly using the configuration and without enumerating any policy: it is easy to compute the number $a(m)$ of trajectories anonymized by node $m$ of $T$, as the difference between the number of trajectories passed up by $m$'s children and the number of trajectories passed up by $m$ itself; $Cost_c$ simply multiplies $a(m)$ with the area of the cloak sequence represented by $m$, summing up over all $m \in T$.

DEFINITION 4. *[**Configuration cost**] Let $U$ be a set of trajectories and $C$ be a configuration of the G-tree $T$. We define the* cost *of $C$* for *$U$, denoted $Cost_c(C, U)$, as*

$$Cost_c(C, U) := \sum_{m \in nodes(T)} f(m, C) \times Cost(m)$$

*where $f(m, C)$ is given by*

$$f(m, C) = \begin{cases} (d(m) - C(m)), & \textit{If m is leaf} \\ ((\sum_{i=1}^{l} C(m_i)) - C(m)), & \textit{If m is internal} \end{cases}$$

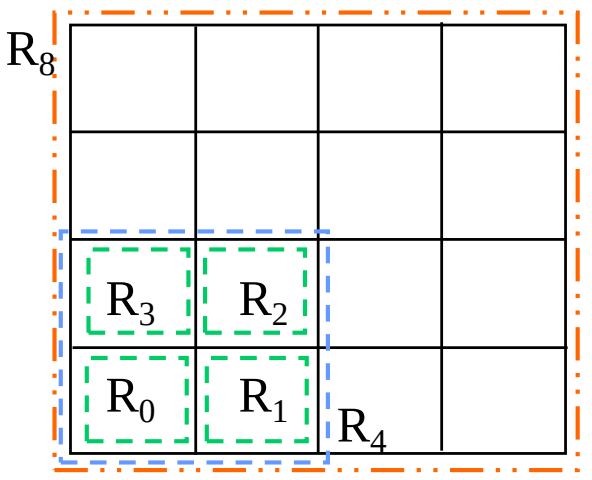

Figure 4: Quad-tree Partitioning

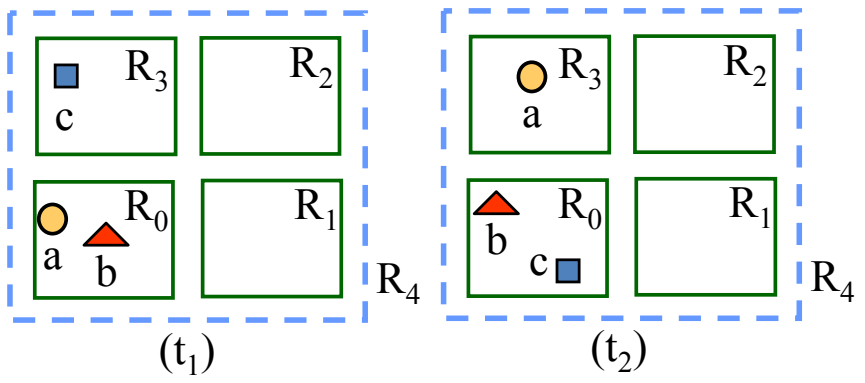

Figure 5: Trajectories of length 2

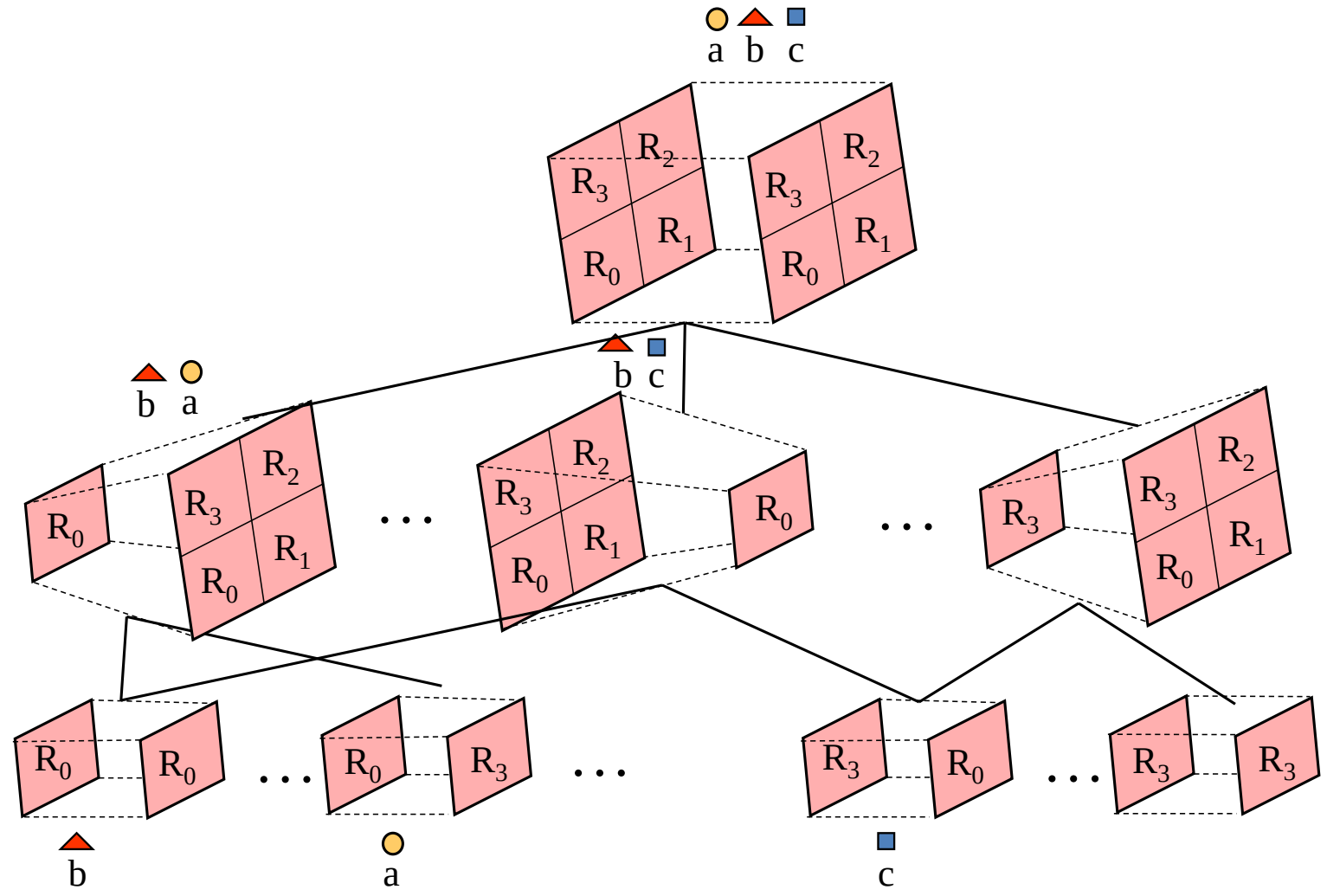

Figure 6: Trajectory Anonymization using Quad-tree

*where $m_1 \dots m_l$ are the children of $m$ and Cost(m) is the sum of areas of the cloaks in the sequence corresponding to node $m$.* □

We can show that the configuration cost is precisely the cost of the represented policies:

(**Extended Version) Lemma** 4. *Given a set $U$ of $n$ trajectories of length l, a G-tree $T$ of quad-cloak sequences of length l, a policy $P$ that usees cloak sequences from $T$ and a configuration $C$ representing $P$'s equivalence class, we have $Cost_c(C, U) = Cost(P, U)$.*

Thus finding the optimum quad-tree policy that uses cloak sequences from $T$ to anonymize a set $U$ of trajectories is equivalent to finding the optimum configuration $C$ of the tree $T$ w.r.t. $U$. Our algorithm does exactly that, i.e. we first find the minimum cost configuration and then materialize a policy corresponding to it.

**Checking Sender Anonymity from Configurations** Since the algorithm manipulates configurations instead of policies, we need a check that a configuration corresponds to *TP*-aware sender k-anonymous policies. By Lemma 1(b), either all represented policies qualify, or none does. It turns out that it suffices to check directly that the configuration satisfies a property we call *k-summing*.

DEFINITION 5. *[**k-summing**] Let $U$ be a set of trajectories and $C$ a configuration of the tree $T$ rooted at $r$. $C$ is a k-summing configuration if*

- *for a leaf node $m$*
  - *(i) if $d(m) < k$, then $C(m) = d(m)$.*
  - *(ii) if $d(m) \geq k$, then either $C(m) = d(m)$ or $C(m) \leq (d(m) - k)$.*
- *for an internal node $m$ let $\Delta = \sum_{i=1}^{f} C(m_i)$, where $m_1 \dots m_f$ are the children of $m$ in $T$*
  - *(iii) if $\Delta < k$, then $C(m) = \Delta$.*
  - *(iv) if $\Delta \geq k$, then either $C(m) = \Delta$ or $C(m) \leq (\Delta - k)$.*

Intuitively, in Definition 5, clause (i) states that if the quad-cloak sequence corresponding to node $m$ masks less than $k$ trajectories, none of them can be anonymized by $m$ lest k-anonymity be compromised. The responsibility of anonymizing all $d(m)$ of them is passed up to $m$'s ancestors ($C(m) = d(m)$). By clause (ii), if there are at least $k$ trajectories, then either all of them are passed up, or at most $d(m) - k$ (since at least $k$ must be anonymized to the same cloak sequence to preserve k-anonymity). For an internal node $m$, $\Delta$ represents the number of trajectories whose anonymization responsibility is passed up from $m$'s children to $m$. If there are too few of them (less than $k$) then they cannot be anonymized using the cloak sequence of $m$, who in turn passes the responsibility to its ancestors (in clause (iii)). Otherwise, $m$ has the choice of either anonymizing none of them ($C(m) = \Delta$ in clause (iv)), or anonymizing at least $k$ and passing up at most $\Delta - k$.

LEMMA 2. *Let $T$ be a G-tree of quad-cloak sequences and $U$ be a set of trajectories. Let $C$ be a configuration of $T$ for anonymizing $U$, and $P$ be a policy in the equivalence class $C$ represents. $P$ provides* TP*-aware k-anonymity to $U$ if and only if $C$ is a k-summing configuration.*

Lemma 2 justifies an algorithm that explores the space of k-summing configurations, in search for a complete minimum-cost configuration. But for a set of $n$ trajectories and a G-tree $T$ with $m$ nodes there are $O(n^m)$ possible configurations. Next we describe the second property of the policies in $\mathcal{P}_T$ that enables a divide-and-conquer approach to find the optimum k-summing configuration.

**Property 2: Optimum cost of anonymizing a subset of trajectories using a node in $T$ can be computed locally** Let $C$ be a k-summing configuration $C$ of a G-tree $T$ of quad-cloak sequences. For a node $m$ in $T$, $C(m)$ represents the number of unanonymized trajectories passed up by $m$. These trajectories are anonymized at one of the ancestors of $m$ and hence they do not affect how the $d(m) - C(m)$ trajectories are anonymized using $m$ and its descendants. Thus for a given value of $C(m)$, one can optimize the anonymization of $d(m) - C(m)$ trajectories using $m$ and its descendants independently of the rest of the trajectories and the rest of $T$. Before we describe how we compute this local optimum for each $m$, we need to point out that at this stage we don't know the value of $C(m)$ in the optimum configuration. For this reason we compute the optimum costs of passing up $0, 1 \dots d(m)$ trajectories at $m$ i.e. all possible values of $C(m)$. For each pair $(m, u)$ such that $C(m) = u$, the minimum cost is computed among all possible configurations of the subtree rooted at $m$ (as there are many possible configurations with $C(m) = u$).

**Computing all local optimum costs** To compute the (local) optimum value of passing up $u$ trajectories at node $m$, the algorithm considers all possible counts $\langle 0, 1 \dots d(m_1)\rangle$, $\langle 0, 1 \dots d(m_2)\rangle, \dots$, $\langle 0, 1, \dots, d(m_f)\rangle$ of trajectories passed by $m$'s children $m_1, \dots, m_f$ respectively. Then it recursively computes the corresponding minimum cost for each $(m_i, u_i)$ pair. Redundant cost re-computation for $m, u$ pairs is avoided by a memoization technique: i.e., by storing the result in the corresponding cell of a bi-dimensional matrix $M$ indexed by the nodes of $T$ and values of $u$. To enable the easy retrieval of the min-cost configuration from $M$, the entries for node $m$ carry, besides the minimum cost, some bookkeeping information relating to the configurations of the children of $m$.

This yields the following dynamic programming algorithm *Traj-anon* that, given a set $U$ of trajectories of length $l$ and a G-tree $T$ with cloak sequences of length $l$, fills in a configuration matrix $M$ of dimension $|T| \times |U|$, where $|T|$ represents the number of nodes in $T$ and $|U|$ the number of trajectories in $U$. Each entry $M[m][u]$ in the matrix is a tuple of the form $\langle x, u_1, u_2, \dots, u_f\rangle$, pertaining to a configuration $C$ such that $C(m) = u$, and where $x$ is the minimum cost of passing up $u$ trajectories, provided that the children $m_1, m_2, \dots, m_f$ of $m$ pass up $u_1, u_2, \dots, u_f$ trajectories respectively. The algorithm traverses the tree $T$ bottom-up starting from the leaf nodes, and for each node and $0 \leq u \leq d(m)$ fills in the entry $M[m][u]$ using the rows from child nodes $m_1, m_2, \dots m_f$.

**Algorithm 1** Traj-anon

1: **for** $1 \leq m \leq |T|$ **do**
2:   **for** $1 \leq u \leq |U|$ **do**
3:     M[m][u] := $\langle \infty, 0, 0, 0, 0\rangle$ {initialize}
4:   **end for**
5: **end for**
6: **for all** node $m \in T$ **do**
7:   **if** (m is a leaf node) and (d(m) < k) **then**
8:     M[m][d(m)] := $\langle 0, 0, 0, 0, 0\rangle$
9:   **else if** (m is a leaf node) and (d(m) $\geq$ k) **then**
10:     M[m][d(m)] := $\langle 0, 0, 0, 0, 0\rangle$
11:     **for** $0 \leq u \leq d(m) - k$ **do**
12:       M[m][u] := $\langle area(m) \times (d(m) - u), 0, 0, 0, 0\rangle$
13:     **end for**
14:   **else** {m is a non-leaf node}
15:     let $m_1, m_2, \dots, m_f$ are children of m
16:     **for all** u in F(m) **do**
17:       pick $u_1 \in F(m_1), u_2 \in F(m_2), \dots, u_f \in F(m_f)$ that minimize the quantity
18:       $x := \sum_{l=1}^{f} M^1[m_l][u_l] + (area(m) \times ((\sum_{l=1}^{f} u_l) - u))$
19:       where $F(m)$ denotes the set $[0..(d(m) - k)] \cup \{d(m)\}$, and $M^1[i][j]$ returns the first component of the tuple at $M[i][j]$
20:       M[m][u] := $\langle x, u_1, u_2, \dots, u_f\rangle$
21:     **end for**
22:   **end if**
23: **end for**
24: **return** M

Function $F(m)$ in line 16 limits the number of trajectories whose anonymization responsibility can be passed up by $m$. Notice that it rules out the values $d(m) - k + 1$ through $d(m) - 1$ since these imply anonymizing less than $k$ trajectories at $m$, which would immediately compromise k-anonymity. Quantity $x$ is the minimum cost among all configurations $C$ with $C(m) = u$ and which satisfies k-summation property. This is computed from the costs of the configurations at the $f$ children, and the number of trajectories anonymized by $m$ $(((\sum_{l=1}^{f} u_l) - u))$ multiplied by $area(m)$ (which denotes the sum of the cloak areas in the sequence represented by $m$). Recall that the cost is the first component of the tuple stored in the matrix entry, whence the need for the projection operation $M^1$.

Notice how the algorithm mirrors Definition 5 to ensure that only k-summing configurations are considered. By Lemma 2, these configurations represent only *TP*-aware sender k-anonymous policies. For instance, line 8 corresponds to case (i) in Definition 5, which prescribes that no trajectories are to be anonymized by $m$ (all $d(m)$ trajectories inside the cloak sequence of $m$ are passed up, $C(m) = d(m)$). Thus by definition of $Cost_c$, the resulting cost is 0, which is what line 8 fills into the first component of $M[m][d(m)]$. Similarly, line 10 gives the cost corresponding to the case in the first disjunct of line (ii) of Definition 5; line 12 corresponds to the second disjunct. We can formally prove that:

LEMMA 3. *Algorithm* **Traj-anon** *computes in each* $M[m][u]$ = $\langle x, u_1, u_2, \dots, u_f\rangle$ *the minimum cost* $x$ *among all k-summing configurations* $C$ *such that* $C(m) = u$ *and* $C(m_i) = u_i$*, with* $m_1, \dots, m_f$ *the children of* m.

**Selecting the optimum configuration** The information in $M$ suffices to retrieve in PTIME a minimum-cost configuration. The optimum configuration is obtained when the optimum cost of $C(r) = 0$ is computed, where $r$ is the root node of $T$ (the root cannot pass up any trajectories as there is no larger cloak sequence to anonymize them with). After that it is easy to retrieve the complete configuration from $M$ in polynomial time by a top-down traversal of $T$. The minimum cost entry $M[r][0]$ for root $r$ lists for its each child $m_i$ the value $C(m_i) = u_i$ leading to the minimum cost. Now inspect for each $m_i$ the $u_i$ entry in $M$, picking again the minimum cost entry for passing up $u_i$ trajectories at $m_i$ and continue recursively until all leaf nodes are reached.

**Complexity analysis** The running time of Algorithm *Traj-anon* is dominated by steps 16-18. For internal node $m$, it ranges each of $u, u_1, u_2, \dots, u_f$ over at most $|U|$ values (since $F(m) \leq d(m) \leq |U|$ for every $m$), resulting in $O(|U|^{f+1})$ iterations where the degree $f$ represents the maximum number of children of a node $m \in T$. Summing up over all nodes $m$ of the tree subspace $T$, we obtain the complexity of *Traj-anon* in $(O|T||U|^{f+1})$. The exponent $f$ of the polynomial depends upon the chosen G-tree $T$ of the G-graph.

**Policy from Configuration** We do not enumerate all the policies of the equivalence class corresponding to the optimum configuration. Note that a configuration $C$ is exponentially more succinct than an explicit listing of the policies it represents; if we focus on any node $m$ alone, there are exponentially many ways to pick $C(m)$ trajectories among those occurring in $m$. Yet, we can obtain one of the policies $C$ represents in linear time by non-deterministically selecting the $C(m)$ trajectories for each node $m$.

### *4.1.2 Choosing the G-tree for l-Approximation*

Our approach for finding an approximation solution to the problem of optimum *TP*-ware sender k-anonymity using quad-tree policy consists of *a)* identifying a subset $S'$ of all the possible quad-cloak sequences and *b)* finding the optimum policy among those policies that only uses the cloak sequences from $S'$.

In Section 4.1.1 we described an algorithm *Traj-anon* that can find the optimum policy w.r.t. any G-tree of the G-graph $G$ of quad-cloak sequences. In this section we show how to choose a G-tree $T_u$ such that the optimum policy w.r.t. $T_u$ is a bounded approximation of the overall optimum policy w.r.t. $G$. $T_u$ is obtained by limiting the choice of cloak sequences to *uniform cloak sequences*.

**Uniform Cloak-Sequence Tree** Let $D$ be a G-graph of quad-cloak sequences of length $l$ that use quadrants of a quad-tree $Q$. Consider a quad-cloak sequence in $D$ in which all cloaks have *the same size*. We call such a cloak sequence *uniform quad-cloak sequence*. The cloak sequences $\langle R_0, R_0 \rangle$, $\langle R_0, R_3 \rangle$ and $\langle R_4, R_4 \rangle$, shown in Figure 6, are examples of uniform quad-cloak sequences. Let $s$ be a uniform quad-cloak sequence in $D$. Let $s_p$ be the cloak sequence obtained by replacing each cloak in $s$ with its parent in $Q$. We refer to $s_p$ as the *total 1-step generalization* of $s$. The subset of uniform quad-cloak sequences from $D$ and the total 1-step generalization function define a tree $T_u$ as follows:

- each uniform quad-cloak sequence in $D$ is a node in $T_u$.
- If $s_p$ is the total 1-step generalization of $s \in T_u$, then $s_p \in T_u$ and we set $s_p$ as the parent of $s$.

It is easy to check that $T_u$ is a G-tree of G-graph $D$.

We refer to this tree as the *Uniform Cloak-Sequence Tree (U-Tree)* since it includes only and all the uniform quad-cloak sequences of the G-graph. The root of the U-tree is the sequence of quad-cloaks corresponding to the root of the quad tree $Q$. The leaf nodes are the uniform cloak sequences where each cloak is a quad-cloak corresponding to a leaf node of $Q$. The intermediate nodes are uniform quad-cloak sequences where each cloak is a quad-cloak corresponding to a non-leaf node of $Q$. The height of the U-tree is the same as that of $Q$ i.e. $h$, since for each leaf uniform cloak sequence $h-1$ successive total 1-step generalizations lead to the root uniform cloak sequence.

EXAMPLE 10. *For the G-graph $G$ shown in Figure 6, the U-tree $T_u$ consists of the nodes on the bottom level (e.g. $\langle R_0, R_0 \rangle$) and the root $\langle R_4, R_4 \rangle$ of $G$ (missing $G$'s second level). The edges in $T_u$ connect all bottom level nodes to the root. Notice that the parent node is obtained using total 1-step generalization of the child nodes in $T_u$. Also the parent $\langle R_4, R_4 \rangle$ and its child nodes are in an ancestor-descendant relationship in $G$. Since the length of the cloak sequences is $l = 2$, each node in $T_u$ has $4^2$ children.* □

**Uniform policy** A policy that only uses uniform quad-cloak sequences in the bundles is referred to as *uniform* quad-cloak policy. Note that in a uniform quad-cloak policy, the cloaks are of the same size within a bundle, but not necessarily across bundles.

THEOREM 2. *Given a set $U$ of trajectories of length $l$, a quad-tree $Q$ and degree of anonymity $k$, the cost of the optimum* uniform *policy that provides* TP*-aware sender $k$-anonymity is at most $l$ times that of the overall optimum policy that provides* TP*-aware sender $k$-anonymity.*

To obtain the optimum uniform policy that anonymizes a set $U$ of trajectories of length $l$, we call the Traj-anon algorithm with U-tree $T_u$ of length $l$ as input. For a U-tree $T_u$ of length $l$, each non-leaf node has $4^l$ child nodes. Substituting this value for $f$ in algorithm Traj-anon, each entry $M[m][u]$ needs to store $4^l$ optimum costs corresponding to $4^l$ child nodes and the number of iterations (Step 17) needed to compute a matrix entry are bounded by $O(|U|^{4^l+1})$. The obtained configuration represents the equivalence class of policies that have the optimum cost among all the uniform policies, for anonymizing $U$. By Lemma 3 and Theorem 2, we have:

THEOREM 3. *When taking the U-tree as input, Algorithm* **Traj-anon** *computes an $l$-approximation solution to the problem of optimum offline $TP$-aware sender $k$-anonymity.*

As described earlier the complexity of Algorithm *Traj-anon* depends upon the maximum degree of a node in the input G-tree. When the input is the U-tree, *Traj-anon* runs in $(O|T_u||U|^{4^l+1})$. Clearly, the exponent $4^l + 1$ is impractically high as we expect a large number of trajectories. In the next section we describe our optimization techniques to reduce the complexity of *Traj-anon* to low-degree PTIME.

### *4.1.3 Optimizations*

In this section we describe optimizations to reduce the complexity of the Traj-anon algorithm without degrading the approximation factor. Due to space limitations, we sketch the high-level ideas, relegating details to Section 4.2.

Recall that the exponent in the complexity of the *Traj-anon* algorithm is determined by the degree (branching factor) $f$ of the input tree, and that in the case of a U-tree, $f = 4^l$ (as each of the $l$ cloaks in a node $n$ is split into 4 sub-quadrants in the children of $n$).

Our first optimization modifies the U-tree $T_u$ (in a strategic way) to guarantee a bounded degree $f = 4$, without eliminating any nodes from $T_u$. This reduces the complexity of finding the optimum configuration of the new tree structure using Traj-anon, without affecting the approximation factor. To this end, we use another type of $G$-tree, called a *US-tree*, that splits "slower" than the U-tree. The idea is to spread the original $4^l$-way split that a U-tree node undergoes in a single level into $l$ US-tree levels of 4-way splits. The first-level node (i.e., the original node) splits only the quadrant at the first snapshot, whereas the four second-level nodes split only the quadrants at the second snapshot, and so on. After $l$ levels in the US-tree, the resulting $4^l$ nodes become uniform quad-cloak sequences again and are exactly the $4^l$ direct children of the original node in the U-tree. Note that the US-tree has a constant degree $f = 4$ and roughly half more nodes than the original U-tree.

The *Traj-anon* algorithm can be applied to the US-tree $T_{us}$, yielding an improved time complexity $O(|T_{us}||U|^5)$ (just substitute 4 for $f$ in the complexity analysis of Section 4.1.1). Furthermore, since all nodes in the original U-tree $T_u$ are included in the corresponding US-tree $T_{us}$, *Traj-anon* is guaranteed to find a policy in $T_{us}$ that is no worse than what it can find in $T_u$. In fact, the optimal policy from a US-tree can have a potentially better cost because there are more nodes (i.e., quad-cloak sequences) to be chosen from.

While the above improvement leads to a polynomial time algorithm with constant exponent 5, this is still impractically high given the typical number of LBS users in a metro city (in the range of 1 million for a city like San Francisco).

We apply a second optimization idea: we adapt our algorithm from quad-tree to binary-tree partitioning of the space, i.e. each quadrant can be split into 2 semi-quadrants, rather than 4 sub-quadrants. We construct $T_{usb}$, the binary tree built from $T_{us}$ above by extending it with nodes obtained by splitting quadrants into 2 semi-quadrants. This immediately lowers the node degree bound from $f = 4$ to $f = 2$, yielding complexity $O(|T_{usb}||U|^3)$.

In the next section we describe a succession of additional optimizations yielding the *Smart Traj-anon* algorithm that has reduced complexity of $O(|T_{usb}|(kh)^2)$, where $h$ is the height of $T_{usb}$ and $k$ is the desired level of anonymity. As an additional optimization, we do not materialize the complete binary semi-quad tree: instead, we split a (semi-)quadrant only if one of its 2 children contains at least $k$ trajectories. Notice that due to this construction $|T_{usb}|$ depends on the spatial distribution of trajectories, but is bounded in the worst case by the number of trajectories $|U|$. Therefore for fixed $k$ and $h$, Smart Traj-anon scales linearly with $|U|$.

## 4.2 Optimizations Details (Extended Version)

### 4.2.1 US-tree

Given a T-uniform tree $T_u$ of uniform cloak sequences of length $l$, we introduce intermediate nodes (cloak sequences of length $l$) between a non-leaf node $m \in T_u$ and it's child nodes such that the resulting structure has the following properties:

- it is a $G$-tree .
- has all the nodes of $T_u$ (and some additional nodes).
- $y$ is an ancestor of $x$ in this $G$-tree, if $y$ is parent of $x$ in $T_u$.
- each non-leaf node in this $G$-tree has exactly 4 child nodes.

We refer to this $G$-tree as *US-tree*. The nodes that are inserted between a node $m \in T$ and its children are not uniform cloak sequences and are obtained by *ordered* 1-step generalization, that we describe next.

**Ordered 1-step generalization.** As described earlier, there are $l$ 1-step generalizations of a cloak sequence of length $l$, one corresponding to each cloak in the cloak sequence. *Ordered 1-step generalization* refers to the process of obtaining $l$ sequence of cloaks by $l$ "successive" 1-step generalizations, such that the $i_{th}$ 1-step generalization is obtained by replacing the $i_{th}$ cloaks in the cloak sequences obtained by $(i-1)_{th}$ 1-step generalization.

Given a T-uniform $T_u$ of length $l$, we obtain the US-tree by inserting intermediate nodes, between node $m$ and its $4^l$ child nodes, that are obtained by ordered 1-step generalizations of the child nodes. For each child node, we obtain $l$ cloak sequences using ordered 1-step generalization and during the computation of ordered 1-step generalization, the cloak sequence obtained by $i_{th}$ 1-step generalization is made a parent of the $(i-1)_{th}$ 1-step generalization. Each each child node, we obtain the node $m$ in the $l^{th}$ 1-step generalization.

We use only one node to represent a cloak sequence even if is obtained via total 1-step generalizations or two or more nodes. For all the child nodes, consider the $1_{st}$ 1-step generalization in the ordered 1-step generalization. Since each quadrant in $Q$ has 4 child nodes, there are 4 child nodes that have identical $1_{st}$ 1-step generalization. Thus each of these intermediate nodes has 4 child nodes. Similarly each cloak sequence obtained in the $i_{th}$ 1-step generalization is common for 4 intermediate nodes that were obtained $(i-1)_{th}$ 1-step generalization. In the $l^{th}$ 1-step generalization we obtained the node $m$ from the 4 intermediate nodes obtained by $(l-1)^{th}$ 1-step generalization.

Thus even after inserting the intermediate nodes, the resulting structure is a US-tree, containing all the nodes of $T_u$, where each non-leaf node has exactly 4 child nodes. Moreover since the parent cloak sequences a 1-step generalization of its child, the parent nodes in $G$-tree completely masks their child nodes.

Next we adapt the Traj-anon algorithm to find the optimum configuration $C$ for for a given US-tree $T_{us}$ and a set $U$ of users. $C$ represents the equivalence class of policies that has the optimum cost among policies that use cloak sequences from $T_{us}$ to anonymize the set $U$ of user-history objects. Moreover the cost of these policies is never worse then the cost of optimum policy that uses only uniform cloak sequences i.e. nodes in $T_u$.

Since $T_{us}$ is a quad-tree the complexity of the above algorithm dominated by the steps 16-18 is O($|T_{us}||U|^5$) where $|T_{us}|$ represents the number of nodes in $T_{us}$and $|U|$ represents the number of user-history objects. Note that ordered 1-step generalization inserts $O(4^l)$ nodes between a non-leaf node and its children in $T_u$, therefore the resulting US-tree $T_{us}$ has $O(m * 4^l)$ more nodes then $T_u$ but this number is independent of the number of trajectories hence

**Algorithm 2** Modified Traj-anon

1: **for** $1 \leq m \leq |T_{us}|$ **do**
2:   **for** $1 \leq u \leq |U|$ **do**
3:     M[m][u] := $\langle \infty, 0, 0, 0, 0 \rangle$ {initialize}
4:   **end for**
5: **end for**
6: **for all** node $m \in T_{us}$ **do**
7:   **if** (m is a leaf node) and (d(m) < k) **then**
8:     M[m][d(m)] := $\langle 0, 0, 0, 0, 0 \rangle$
9:   **else if** (m is a leaf node) and (d(m) $\geq$ k) **then**
10:     M[m][d(m)] := $\langle 0, 0, 0, 0, 0 \rangle$
11:     **for** $0 \leq u \leq d(m) - k$ **do**
12:       M[m][u] := $\langle area(m) \times (d(m) - u), 0, 0, 0, 0 \rangle$
13:     **end for**
14:   **else** {m is a non-leaf node}
15:     let $m_1, m_2, \cdots, m_4$ are children of m
16:     **for all** u in F(m) **do**
17:       pick $u_1 \in F(m_1), u_2 \in F(m_2), \cdots, u_4 \in F(m_4)$ that minimize the quantity
18:       $x := \sum_{l=1}^{4} M^1[m_l][u_l]$ + $(area(m) \times ((\sum_{l=1}^{4} u_l) - u))$
19:       where $F(m)$ denotes the set $[0..(d(m) - k)] \cup \{d(m)\}$, and $M^1[i][j]$ returns the first component of the tuple at $M[i][j]$
20:       M[m][u] := $\langle x, u_1, u_2, \cdots, u_4 \rangle$
21:     **end for**
22:   **end if**
23: **end for**
24: **return** M

a constant factor. The reduced complexity results in reduced running time for finding the optimum configuration as observed in our experiments.

Even though Traj-anon for a US-tree of length $l$ has reduced complexity in comparison to Traj-anon for a T-uniform of length $l$, we do trade cost to achieve better complexity. This is shown by the following result.

(**Extended Version) Lemma** 5. *Given a US-tree $T_{us}$ of length l, T-uniform $T_u$ of length l and a set U of user-history objects of length l and the level of anonymity k, the cost of the optimum k-summing configuration for $T_{us}$ is never more than the cost of the optimum k-summation configuration for $T_u$.*

As a result the upper bound on the approximation ration i.e $l$ still holds for a US-tree. This follows directly from Lemma 5.

(**Extended Version) Lemma** 6. *Algorithm Modified Traj-anon computes the l-approximation solution to the problem of optimum offline TP-aware sender k-anonymity.*

Moreover the average case cost of optimum policy that uses $T_{us}$ is lower than the policy that uses $T_u$ since when using $T_{us}$ a policy has more options for choosing a sequence of cloaks to anonymize a user object.

While the above optimization leads to polynomial time algorithm with constant exponent, the degree 5 is still high given the typical number of LBS users in a metro city (in the range of 1 million for city like San Francisco). Next we describe a pair of optimizations to further reduce the complexity of Traj-anon and run-time optimizations to achieve practical running time, while guaranteeing to preserve the approximation bound. These optimizations are similar in spirit to the optimizations for quad-tree based snapshot policy-aware sender k-anonymity described in [16] with the difference that the tree in our study consists of a quad-tree of quad-cloak sequences.

### 4.2.2 From Quad to Binary Tree

In the US-tree, if anonymizing a trajectory to a node does not provide the desired k-anonymity, the next possible option is the parent node. Since the parent node is 1-step generalization of the child, the cost of the new cloak in the parent cloak sequence is 4 times that of the replaced cloak in the child cloak sequence. As shown in [24] and [16], the granularity of this cost increase can be reduced by converting a quad-tree into a binary tree by using *semi-quadrants* as cloaks (where a semi-quadrant is obtained by splitting a quadrant into two rectangles, either vertically or horizontally). Use semi-quadrants in the uniform cloak sequences to anonymize the trajectories leads to following:

- The Traj-anon algorithm finds the optimum policy among all the policies that use uniform semi-quadrant cloak sequences.
- The policy obtained using Traj-anon is $l$-approximation of the optimum policy that use semi-quadrant cloak sequences.
- Using total 1-step generalization we can obtain binary US-tree that contains all the uniform semi-quadrant cloak sequences, and in which each node has exactly 2 child nodes.

As a result of this optimization, the complexity of the optimized Traj-anon for a binary US-tree $T_{usb}$ and a set of trajectories $U$, is O($|T_{usb}||U|^3$). In addition, since the semi-quadrants are smaller than the quadrants, this optimization also reduces the average cost of anonymization.

### 4.2.3 Pruning Suboptimal Configurations

For any node *m* of the USeq-Btree, in the for loop of step 16, Traj-anon inspects $(d(m) - k + 1)$ configurations (all possible k-summing configurations). We realize that some of these configurations need not be considered, as they are guaranteed to be sub-optimal. In fact we claim the following lemma:

(**Extended Version) Lemma** 7. *For a node $m$ with height $h(m)$ (where the height of the root is* 0*), any configuration in which $m$ passes up to its ancestors the cloaking responsibility of more than $k(h(m) + 1)$ but less than $d(m)$ trajectories, is not optimal.*

By Lemma 7, it suffices to compute $k(h(m)+1)$ configurations, by simply replacing function $F$ in step 16 of algorithm Modified Traj-anon with function $F'(m) = [0..(k(h(m) + 1))] \cup \{d(m)\}$. Thus for a non-leaf node $m$, the algorithm computes $O(kh)$ configurations and to compute each such configuration, the "pick" action iterates over $O(kh)$ configurations of $m$'s two children. This leads to a new upper bound of the overall running time, $O(|T_{usb}|(kh)^3)$.

### 4.2.4 Precomputation

Similar to $Bulk_{dp}$, there is significant overlap in the computations across iterations of For loop in Step 16 of Modified Traj-anon. For example, if one iteration works on the $M$ entry for $(m, u)$, inspecting for instance $(m_1, u_1)$ and $(m_2, u_2)$ such that $u_1 + u_2 = u$, then the next iteration $(m, u + 1)$ will inspect the cases $(m_1, u_1 + 1)$, $(m_2, u_2)$ and $(m_1, u_1)$, $(m_2, u_2 + 1)$, among others. The idea is to reuse this computation across iterations.

To this end, we stage the computation in 2 parts. In the first stage we iterate over the $O(kh)$ configurations of both children to compute a temporary matrix $temp$. There are $O(kh)$ entries in this matrix and the complexity of this stage is bounded by $O((kh)^2)$. In the second stage, we create $O(kh)$ configurations using the $O(kh)$ entries of temp. Thus the running time for the second stage is also bounded by $O((kh)^2)$. Therefore the overall complexity of the modified step 16 is $O((kh)^2)$ and the overall complexity of the modified algorithm becomes $O(|T_{usb}|(kh)^2)$.

## 4.3 Runtime Pruning

We implement a runtime optimization to further reduce the running time of the modified Traj-anon. We create the binary US-tree top-down by successively splitting the semi-quadrants, starting from the root node. But we do not eagerly materialize all nodes of the binary US-tree, instead, we split a (semi-)quadrant only if it contains sufficient users to maintain anonymity.

# 5. EXPERIMENTS

In this section we describe a set of experiments to evaluate the effectiveness of our Smart (optimized) Traj-anon algorithm. We evaluate scalability and performance and compare the cost of anonymization and execution time of Smart Traj-anon with a set of alternate anonymization techniques. Since we focus exclusively on the Smart version of Traj-anon, we will drop the qualifier in the remainder of the section.

Our experiments show that Traj-anon scales linearly with the number of trajectories and can anonymize up to *2 million* trajectories of length 30 within *4 min*. We show that the other anonymization techniques either have higher anonymization cost (up to *100 times* higher) or running time (up to *2000 times* slower).

**Trajectory Data.** Due to legal hurdles we could not obtain actual user trajectory data from LBS providers, but we were able to resort to the *Brinkhoff generator* [14] to generate the trajectory data for our experiments. The Brinkhoff generator has been widely used to generate moving object data for studies in various fields, including location-based services and beyond. It takes as input the road network of a metro area and generates trajectories of various classes of moving objects that are constrained by the road network. The classes differ in number and speed with which the trajectories move relative to each other (e.g. cars, bikes, pedestrians, etc.). We generated a master data set of 2 million trajectories of length 30, with 5 different classes of moving objects, using the actual road network of the San Francisco Bay area. Then we drew random samples of increasing number of trajectories (10k, 50k, 100k etc.) of length 10 and 30.

**Platform.** Unless otherwise stated all our experiments run on a Linux server with an Intel Xeon Processor (2.8GHz) and 32G memory. In one experiment, we had to use a machine with Intel Pentium Core2 Duo processor (2.4Ghz) with 2 GB RAM and running Cygwin on Windows XP because the binary we got from the authors of [30] was compiled for that configuration.

**Anonymity degree.** In all experiments, $k = 50$.

## 5.1 Scalability

In the first set of experiments we evaluate the scalability of the Traj-anon algorithm by increasing the number of trajectories to be anonymized, from 10k to 2 million, for a fixed k=50 (we consider both trajectory length 10 and 30). As shown in Figure 9, the algorithm scales linearly with the number of trajectories. In particular, Traj-anon anonymizes 2 million trajectories of length 30 in less than 4 min. Figure 10 breaks down the running time into a) loading the user trajectories from a file to the main memory data structures, b) obtaining the optimum configuration for the user trajectories and, c) obtaining the policy from the configuration (as expected this time is negligible, under 1%).

## 5.2 Related anonymization techniques

We are unaware of any competing TP-aware sender anonymity solutions. As detailed in Section 6, the previously proposed algorithms for trajectory-aware sender k-anonymity do not defend against policy-aware attackers, and as shown in Section 1 policy-aware snapshot sender k-anonymizing algorithms [16] do not defend against trajectory-aware attacks. Since we couldn't find direct competitors, we created some by leveraging existing work.

As a baseline approach we decided to extend an algorithm for policy-aware snapshot sender k-anonymity to *TP*-aware sender k-anonymity. We chose the $Bulk_{dp}$ algorithm in [16] since it provides the optimum anonymization for a snapshot and uses (semi-)-quadrant cloaks (just like Traj-anon).

We also considered solutions proposed for *trajectory anonymity*, a privacy problem orthogonal to sender anonymity. In trajectory anonymization, the goal is to anonymize user trajectories such that an attacker, who knows locations of users in certain snapshots (partial trajectories), cannot infer whether a user's trajectory passes through a particular location. Trajectory anonymity tries to hide the user's whereabouts, while sender anonymity assumes them as known and focuses instead on hiding the identity of request senders. Due to the different goals and assumptions of the two privacy guarantees, some of the data transformation techniques employed in trajectory anonymization (such as *deletion of locations*, *addition of locations*, and *shifting locations* from a trajectory), do not apply to sender anonymity. Despite the differences, we identified a class of trajectory anonymization solutions whose techniques can in principle be adapted to provider *TP*-aware sender k-anonymity. This class of solutions use some clustering algorithm to partition user trajectories into groups of k trajectories and then applies other data transformations (described above) to preserve trajectory anonymity. We realized that one can adapt the clustering techniques to obtain the bundles used in offline *TP*-aware sender k-anonymization. We borrowed the clustering techniques from state-of-the-art trajectory anonymization solutions [25, 30] to obtain three different competing solutions for offline *TP*-aware sender k-anonymity.

Next we describe these three solutions along with the baseline approach based on snapshot P-aware sender k-anonymity.

**Baseline TP-aware** is based on the P-aware snapshot anonymization algorithm $Bulk_{dp}$ described in [16]. We format the input trajectory data as a sequence of snapshots. We anonymize the first snapshot of the input trajectory data using $Bulk_{dp}$ and group together the trajectories whose locations in the first snapshot are anonymized to the same cloak. Since $Bulk_{dp}$ provides policy-aware sender k-anonymity each group must have at least $k$ members. For each group and for each snapshot, we find the smallest quadrant that masks the locations of the trajectories in the group. Thus for each group we obtain a sequence of quadrants that plays the role of a bundle in the sense of Traj-anon. This anonymization provides *TP*-aware sender k-anonymity since there are at least k trajectories that are anonymized to the same sequence of cloaks.

**Fast Clustering** is based on the *fast TGA* algorithm in [25]. It creates a cluster of k trajectories by first randomly selecting an unanonymized trajectory as the center of the cluster and then adding its k-1 nearest neighbor trajectories to the cluster. The distance between two trajectories is the sum of the "distances" between their locations in each snapshot and the distance between two locations is the logarithm of the area of the smallest axis-parallel minimum bounding rectangle (rectangle whose sides are parallel to the x and y axis of a 2-dimensional plane) that masks the two locations.

**Slow Clustering** is based on the *multi TGA* algorithm in [25]. To create a cluster of k trajectories, it first randomly selects an unanonymized trajectory as the center of the cluster and adds k-1 additional trajectories one by one so as to minimize the cost of the cluster. The cost of a cluster is the sum of the logarithms of areas of axis-parallel MBRs, that masks the locations of the trajectories in the cluster, in each snapshot.

**Hilbert-based Clustering** [30] uses an embedding of two-dimensional into one-dimensional space, associating to each location a Hilbert index which is then used to simplify the nearest-neighbor computation used to cluster trajectories. The original approach requires identification of certain locations in a trajectory as quasi-identifiers (uniquely identifying the user). Since we assume that the entire user trajectory is accessible to the attacker, every location in his trajectory is a potential quasi-identifier. Thus in the input to the Hilbert-based clustering algorithm we specify every location of a trajectory as quasi-identifiers. As a result, the distance between two trajectories is the sum of the absolute difference between the Hilbert indexes of the locations in each snapshot. Since the algorithm computes the clusters of k-1 nearest neighbors for each trajectory independently, two clusters can have some trajectories in common. Clusters with common trajectories are merged.

In the three clustering-based approaches, after computing the clusters, all trajectories in a cluster are anonymized using the sequence of axis-parallel Minimum Bounding Rectangles (MBR) (rectangles whose sides are parallel to the x and y axis) that masks each snapshot of the clustered trajectories. We compare the execution time and cost of anonymization obtained using the four algorithms described above with those of Traj-anon. To make a fair comparison, we modify the output of Traj-anon and replace the quadrants in the cloak sequence with axis-parallel MBRs ensuring that the MBR that replaces a quadrant must be included in the quadrant.

We implemented the Baseline, Fast and Slow Clustering algorithms in C++ and obtained the Hilbert-based Clustering binaries from the authors of [30] (compiled for Cygwin on Windows XP).

Figure 11 shows the cost of anonymizing an increasing number of trajectories (10k, 50k, 100k, 200k,600k and 1M) of length 30, using Traj-anon and the clustering-based algorithms described above (results for length 10 are similar and not shown here; we stopped at 1M trajectories as the other algorithms did not scale).

**Comparison with snapshot-based Baseline.** As shown in Figure 11 the Baseline approach has the highest cost among all the anonymization algorithms. It is significantly more than Traj-anon. For 600k trajectories and more the cost of anonymization with Baseline is *100 times* that of Traj-anon. This is because having optimum cost for one snapshot leads to bigger cost for other snapshots when the trajectories diverge (since the trajectories in a group must be anonymized together in all the snapshots).

**Comparison with Fast Clustering.** As shown in Figure 11 the cost of anonymizing trajectories using fast clustering is more than that with Traj-anon and slow clustering. The difference between the anonymization cost increases with the number of trajectories and for 1 million trajectories of length 30, the anonymization cost of fast clustering is *100 times* more than that of Traj-anon. In terms of execution time, as shown in Figure 12, the fast clustering takes significantly longer in comparison with Traj-anon. For e.g. Traj-anon takes less than *1.5 min* to anonymize 1 million trajectories of length 30 in comparison to *370 min* by fast clustering.

**Comparison with Slow Clustering.** As shown in Figure 11 the cost of anonymizing trajectories using slow clustering is better than that with Traj-anon, by a factor of roughly 4: for 600K (1M) trajectories, Traj-anon obtains a cost of $20 \times 10^{16}(28.5 \times 10^{16})$, Slow clustering a cost of $5 \times 10^{16}(6.6 \times 10^{16})$. But as shown in Figure 12 Slow Clustering is the slowest of all anonymization techniques. It takes over *3 days* for the slow clustering algorithm to anonymize 1 million trajectories of length 30, in comparison under

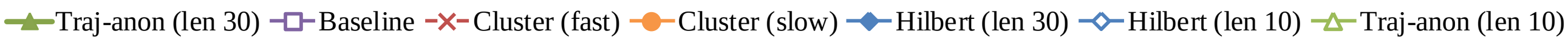

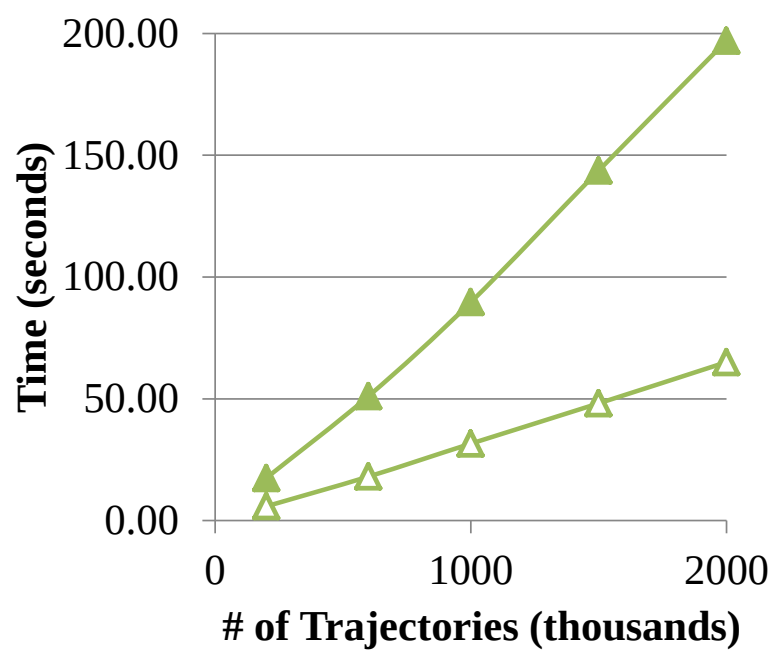

Figure 9: Execution time

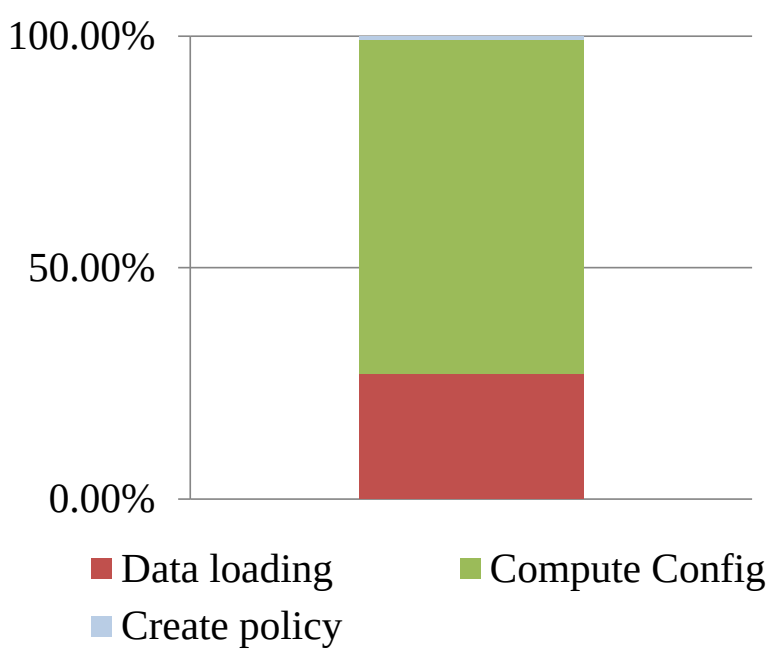

Figure 10: Time spent in various phases

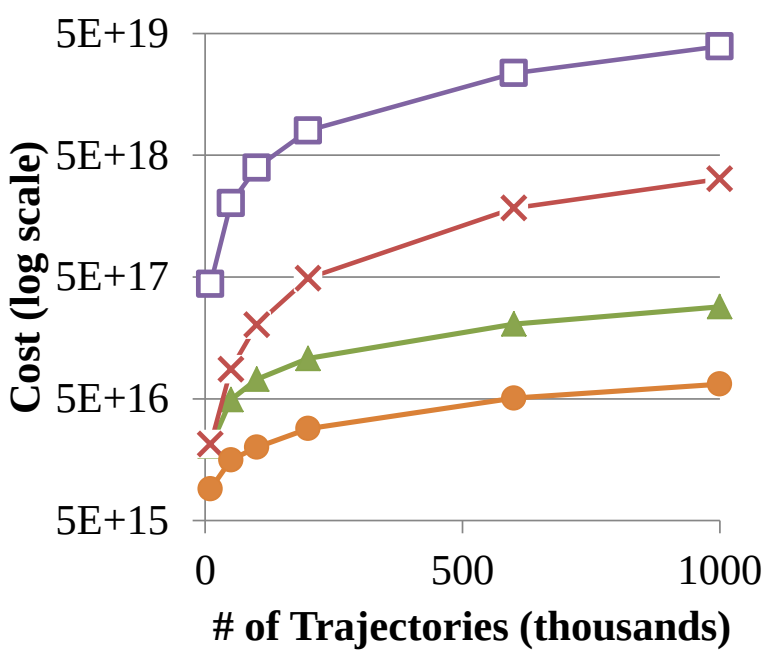

Figure 11: Cost: Clustering vs Traj-anon

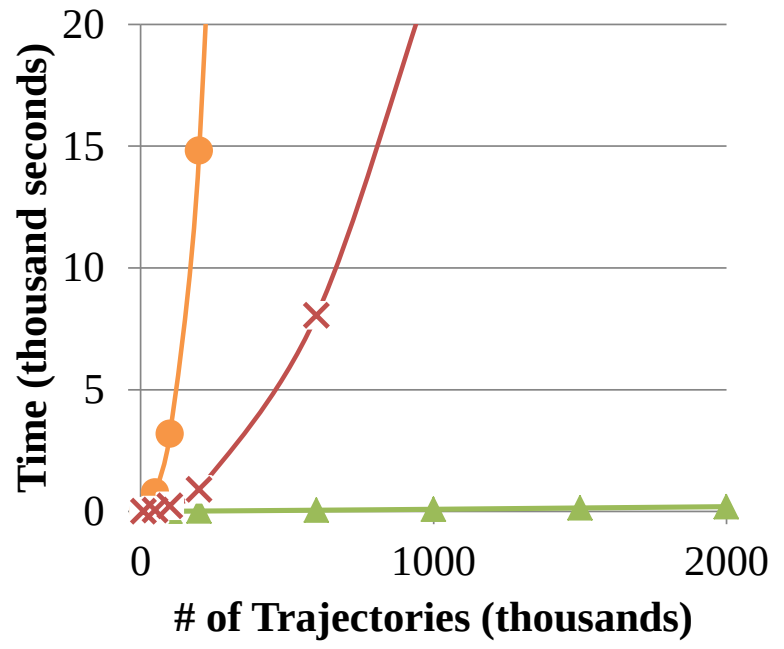

Figure 12: Exec time: Clustering vs Traj-anon

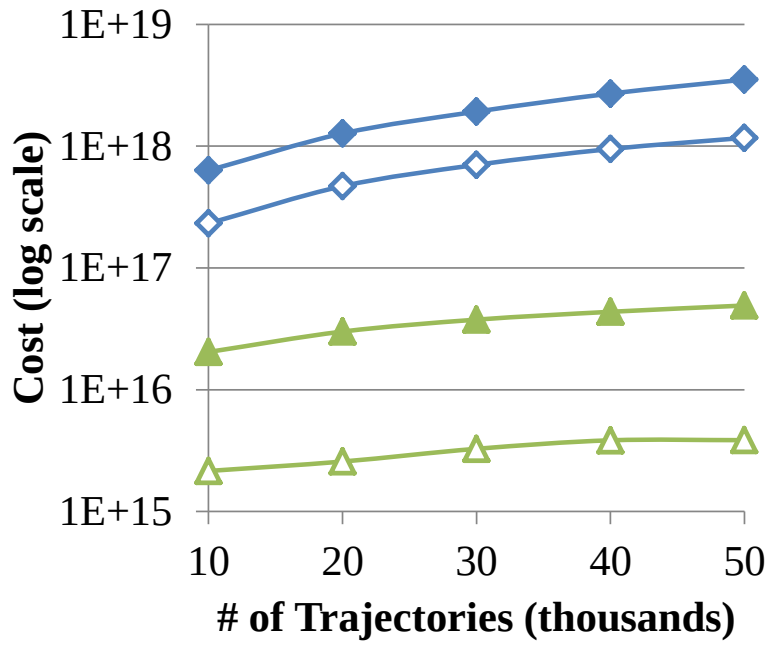

Figure 13: Cost: Hilbert vs Traj-anon

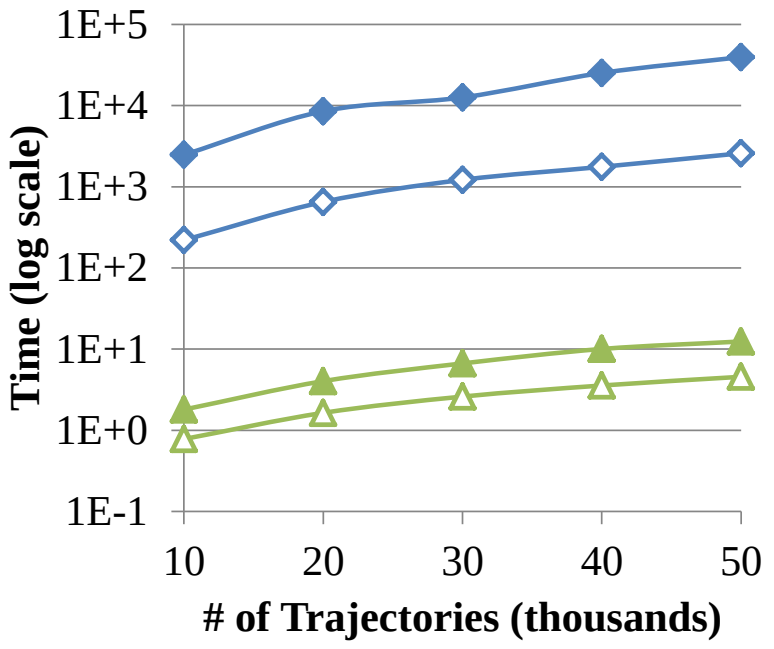

Figure 14: Exec time: Hilbert vs Traj-anon

*1.5 min* with Traj-anon. The poor performance of Slow Clustering is not just accidental (for this data set) but intrinsic to the algorithm due to $O(n^2)$ distance computations between $n$ trajectories.

**Comparison with Hilbert-based Clustering.** Figure 13 compares the anonymization cost of Hilbert-based clustering with that of Traj-anon. We could not process more than 50k trajectories because the Hilbert-based clustering implementation does not scale beyond. The cost of Hilbert-based clustering is considerably higher than Traj-anon. Even for 10k trajectories the cost of Hilbert-based clustering is *30 times* higher than that of Traj-anon and, reaching *50 times* for 40k trajectories. A possible reason for the higher cost of Hilbert-based clustering is the distortion introduced in mapping the 2-dimensional space to a single dimension.

As shown in Figure 14 the Hilbert-based clustering is considerably slower than Traj-anon. It takes over *11 hours* to anonymize 50k trajectories of length 30 using Hilbert-based clustering in comparison to *12 sec* by *Traj-anon*. Even though the Hilbert-based clustering uses a simpler distance function, it is slower due to the high number of comparisons to find the $k$ nearest neighbors.

## 6. RELATED WORK

In the context of LBS, the two aspects of privacy that have received most attention are *trajectory anonymity* and *sender anonymity*.

**Trajectory anonymity.** As detailed in Section 5.2, the line of work on *trajectory anonymity* [25, 30, 28, 22] is complementary to ours: its goal is to hide the user's precise location over a period of time (one is not required to hide the identity of the user), while sender anonymity hides the identity of the user, assuming that the trajectory data falls in the attacker's hand. Even though the problem of trajectory anonymity is orthogonal to the problem studied in this paper, as described in Section 5 a class of clustering based solutions can be adapted to provide offline TP-aware sender k-anonymity, and we have compared against them in detail.

**Classes of Attackers.** The solutions for sender k-anonymity in the context of location-based services can be classified into four categories based on the class of attackers they prevent against:

*Policy-unaware trajectory-unaware:* The solutions [19, 24, 20] in this class are also known as *k-inside* policies [16] as these solutions use tightest cloak (of a pre-defined shape) that includes the sender and k-1 other users. This class of solutions neither preserve privacy against a policy-aware attacker (as shown in [16]) nor against a trajectory-aware attacker (also shown in [12, 29, 15]).

*Policy-aware trajectory-unaware:* This class of solutions [11, 16] ensures that there are at least k users anonymized using the same cloak. The privacy guarantee of these solutions is strictly stronger than the policy-unaware solutions i.e. they also defend against policy-unaware trajectory-unaware attackers but not conversely (for a formal proof see [16]). But as shown in Section 1 they fail to preserve privacy against a policy-aware attacker who is also trajectory-aware.

*Policy-unaware trajectory-aware:* This class of solutions [29, 18, 15, 12] targets anonymity against the trajectory-aware attackers using a sequence of cloaks that masks the user and the same k-1 users for the entire duration of the user trajectory. [15] claims policy-awareness as well, but the claim needs qualification, as it isn't clear what the attacker knows: a) the mapping from a given set of user trajectories to the sequence of cloaks, or b) the algorithm producing this mapping in addition to the mapping itself (this is our sense of policy-awareness). We claim that [15] defends against attacker class a) but not b), and thus gives a weaker guarantee than the one we target here. We illustrate how the *2-sharing* property of [15] allows policy-aware attackers to breach privacy.

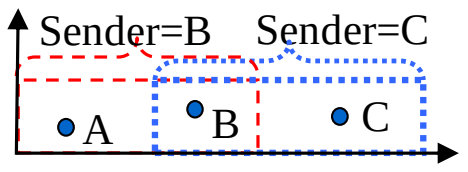

Figure 15: 2-sharing policy

EXAMPLE 11. *Consider the cloaking algorithm in [15] that takes into account the requesting location to generate cloaking groups (set of locations that are cloaked to the same region). For the locations in Figure 15, if the first request is made by $C$ the algorithm groups $C$ with $B$ whereas if the first request is made by $B$ then $B$ and $A$ end up in the same cloaking group to satisfy the 2-sharing property. In the case when the first anonymized request contains the cloak corresponding to $\{C, B\}$, a policy-aware attacker immediately infers that the sender is $C$.* □

*Policy-aware trajectory-aware:* We are unaware of any work that provides policy-aware and trajectory-aware sender k-anonymity and therefore we propose the guarantee in this paper. As illustrated in Section 1 even this privacy guarantee does not allow to completely publish the linkage between multiple requests sent by the same user. It does allow to publish the requests made along a trajectory bundle.

**Trusted LBS.** In the model used in this paper we assume that the LBS provider is a trusted entity and responsible for anonymizing the user requests that it collects over a period of time. We share this assumption with a line of work on trajectory anonymization [25, 30, 28, 10, 23, 22] where the location provider (who logs user trajectories) is trusted and is responsible for anonymization.

**Online vs Offline.** Another contrasting feature between previous trajectory-aware sender anonymity proposals and the one in this paper is the *mode* of anonymization. In [29, 18, 15, 12] LBS requests are anonymized as they are issued i.e. *online* while we anonymize the request log i.e. *offline*. One can possibly use the online solutions for the offline TP-aware sender anonymization but with a necessarily sub-optimal cost since the future movement of the users is not known by the online anonymizer. The cloak that masks a group of $k$ users can become arbitrarily large if their trajectories diverge.

**Beyond sender k-anonymity: l-diversity.** In the setting of relational table anonymization, k-anonymity is viewed as a classical baseline, recently subsumed by stronger guarantees ranging from l-diversity [21] to differential privacy. For the LBS context, this raises the natural question of analogous guarantees that subsume sender k-anonymity. We note that LBS sender privacy is a much younger field, in which even such a fundamental guarantee as sender k-anonymity (especially in its TP-aware form) hadn't been solved until now. We also note that in an LBS context, sender k-anomymity is not weaker than sender l-diversity, actually coinciding with it. To see why, consider an analogy to the "homogeneity attack" that breaks classical k-anoymity but is foiled by l-diversity [21]: there is a possibility that all user histories masked by a bundle send identical requests in a particular snapshot (possibly from different locations). But since a bundle associates a set of requests (no duplicates) with a cloak, all the identical requests are represented by a single request, thus precluding the homogeneity attack, and in fact any attack based on the distribution of request values.

# 7. CONCLUSIONS

We introduce and study the problem of offline trajectory- and policy-aware sender k-anonymity. We show that prior results for snapshot k-anonymity do not apply and that trajectory-awareness leads to strictly stronger attackers, calling for a stronger privacy guarantee. We show that optimum TP-aware anonymization is computationally harder than snapshot P-aware anonymization (NP-complete vs. PTIME). We propose a PTIME $l$-approximation algorithm for trajectories of length $l$ and empirically show its effectiveness.

# APPENDIX

# A. PROOFS

## A.1 Lemma 1

PROOF. **(a)** Let $U$ be a set of $n$ trajectories and policies $P_1$ and $P_2$ are equivalent for anonymizing $U$ w.r.t. a $G$-tree $T$. We describe the cost of anonymizing $U$ using $P_1$ as:

$$Cost(P_1, U) = Cost(m_1) + Cost(m_2) + \ldots + Cost(m_n)$$

where $m_i = P_1(U, u_i) \in T$ for $1 \leq i \leq n$. Note, for $i \neq j$, $m_i$ and $m_j$ can be the same node in $T$. Similarly we describe the cost of anonymizing $U$ using $P_2$ as:

$$Cost(P_2, U) = Cost(m'_1) + Cost(m'_2) + \ldots + Cost(m'_n)$$

where $m'_i = P_2(U, u_i) \in T$ for $1 \leq i \leq n$. Note, for $i \neq j$, $m'_i$ and $m'_j$ can be the same node in $T$. Since $P_1$ and $P_2$ are equivalent, if a node $m \in T$ is used by $P_1$ to anonymize $x$ trajectories in $U$ then $m$ is also used by $P_2$ to anonymize the same number of trajectories in $U$. Therefore,

$$\begin{aligned} &Cost(m_1) + Cost(m_2) + \ldots + Cost(m_n) \\ &\quad = Cost(m'_1) + Cost(m'_2) + \ldots + Cost(m'_n) \end{aligned} \quad (1)$$

because each quadrant appears same number of times on both sides of Equation 1. Hence we have:

$$Cost(P_1, U) = Cost(P_2, U)$$

**(b)** Suppose $P_1$ provide $TP$-aware sender k-anonymity to $U$ w.r.t. $T$. Therefore, for each node $m \in T$, $P_1$ either anonymizes none or at least $k$ trajectories using $m$. We are given that $P_1$ and $P_2$ are equivalent for $T$, therefore they both anonymize the same number of trajectories using $m$. Therefore, $P_2$ either anonymizes none or at least $k$ trajectories using $m$. Thus, $P_2$ also provides $TP$-aware sender k-anonymity to $U$ w.r.t. $T$. Similarly we can show that if $P_2$ provides $TP$-aware sender k-aonymity to $U$, so does $P_1$.

□

## A.2 Lemma 2

PROOF. Let $U$ be a set of trajectories and $T$ be a $G$-tree of quad-cloak sequences. Let $P$ be a policy that uses the cloak sequences from $T$ and $C$ be the configuration representing the class of policies equivalent to $P$.

First we assume that $P$ provides *TP*-aware sender k-anonymity and show that $C$ is k-summing configuration. Since $P$ is *TP*-aware sender k-anonymous, each quad-cloak sequence in $T$ is used in $P$ to anonymize either none or at least $k$ trajectories. Thus

- For a leaf node $m \in T$
  - (i) If $d(m) < k$, then $P$ cannot anonymize any trajectory using $m$, therefore $C(m) = d(m)$.
  - (ii) if $d(m) \geq k$, then $P$ could either anonymize at least $k$ trajectories or *none*. In former case $C(m) \leq (d(m) - k)$ while in later case $C(m) = d(m)$.
- For an internal node $m \in T$, let $\Delta = \sum_{i=1}^{l} C(m_i)$, where $m_1 \ldots m_l$ are the children of $m$ in $T$
  - (iii) if $\Delta < k$ then there are less than $k$ trajectories passed up by children of $m$. Thus $P$ cannot anonymize any trajectory using $m$ and therefore we have $C(m) = \Delta$.
  - (iv) if $\Delta \geq k$ then the children of $m$ passes up at least $k$ trajectories. Therefore, $P$ could either anonymize at least $k$ trajectories or no trajectory using $m$. In former case $C(m) \leq (\Delta - k)$ while in later case $C(m) = \Delta$.

Thus $C$ satisfies k-summing property.

Next we assume that $C$ is k-summing configuration and show that $P$ provides *TP*-aware sender k-anonymity to $U$. Equivalently we show that under $C$, each cloak of $T$ is used to anonymize either none or at least $k$ trajectories. Since $C$ is a k-summing configuration, it implies that:

- for a leaf node $m \in T$
  - (i) if $d(m) < k$, then $C(m) = d(m)$. Thus $P$ does not anonymize any trajectory using $m$.
  - (ii) if $d(m) \geq k$, then either $C(m) = d(m)$ or $C(m) \leq (d(m) - k)$. In the later case $P$ anonymizes at least $k$ trajectory using $m$, while in former case none trajectory at all.
- for an internal node $m \in T$ let $\Delta = \sum_{i=1}^{l} C(m_i)$, where $m_1 \ldots m_l$ are the children of $m$ in $T$
  - (iii) if $\Delta < k$, then $C(m) = \Delta$. Thus $P$ does not anonymize any trajectory using $m$.
  - (iv) if $\Delta \geq k$, then either $C(m) = \Delta$ or $C(m) \leq (\Delta - k)$. In the later case $P$ anonymizes at least $k$ trajectories using $m$, while in former case none trajectories.

Therefore $P$ provides *TP*-aware sender k-anonymity.

□

## A.3 Lemma 3

PROOF. The intuition behind Lemma 3 is that Traj-anon algorithm exhausts the search space of all potential optimal k-summing configurations by utilizing Property 2. To prove it formally, we use structural induction to show that for each node $m$ in the $G$-tree $T$ and an integer $l$ such that $l \leq d(m)$, we have

$$cost_{alg}(m, l) = cost_{min}(cset(m, l))$$

where $cost_{alg}(m, l)$ represents the cost computed by Traj-anon for passing up $l$ (unanonymized) trajectories at $m$ and $cost_{min}(cset(m, l))$ represents minimum cost of passing up $l$ trajectories at $m$ among all such k-summing configurations.

**Basis:** For a leaf node $m$ and an integer $l \leq d(m)$, it is obvious by construction that $cost_{alg}(m, l) = cost_{min}(cset(m, l))$.

**Induction:** Let $m$ be a non-leaf node in the $G$-tree $T$ and $m_1, m_2, \cdots m_f$ be the children of $m$. Let $l$ be an integer such that $l \leq d(m)$ and $cset(m, l)$ be the set of k-summing configuration that passes up $l$ (unanonymized) trajectories at node $m$. We show that for each configuration $g \in cset(m, l)$, $cost_{alg}(m, l) \leq cost(g(m))$, where $cost(g(m))$ represents the cost of $g$ at node $m$. If $g(m_1) = l_1, g(m_2) = l_2, \cdots$ and $g(m_f) = l_f$, the cost of $m$ in $g$ can be written as

$$\begin{aligned} cost(g(m)) := [&cost(g(m_1)) + cost(g(m_2)) + \cdots + cost(g(m_f)) \\ &+ cost(m) \times (l_1 + l_2 + \cdots + l_f - l)] \end{aligned}$$

By induction hypothesis we assume that $cost_{alg}(m_1, l_1) \leq cost(g(m_1))$, and similarly $cost_{alg}(m_2, l_2) \leq cost(g(m_2)), \cdots$, $cost_{alg}(m_i, l_i) \leq cost(g(m_i)), \cdots$,
$cost_{alg}(m_f, l_f) \leq cost(g(m_f))$. Therefore

$$\begin{aligned} &cost_{alg}(m_1, l_1) + cost_{alg}(m_2, l_2) + \cdots + cost_{alg}(m_f, l_f) \\ &\leq cost(g(m_1)) + cost(g(m_2)) + \cdots + cost(g(m_f)) \end{aligned}$$

And by adding the constant value $cost(m) \times (l_1 + l_2 + \cdots + l_f - l)$ to both the sides we get

$$cost_{alg}(m, l) \leq cost(g(m))$$

Similarly, for each node $m$ and each integer $l \leq d(m)$, and each configuration $g \in cset(m, l)$, we can show that $cost_{alg}(m, l) \leq cost(g(m))$. Therefore $cost_{alg}(m, l) = cost_{min}(cset(m, l))$. □

## A.4 Lemma 4

PROOF. We describe the cost of anonymizing the set $U$ of $n$ trajectories using policy $P$ as:

$$\begin{aligned} Cost(P, U) &= \sum_{u \in U} Cost(P(U, u)) \\ &= Cost(m_1) + Cost(m_2) + \ldots + Cost(m_n) \end{aligned} \tag{2}$$

where $m_i = P(U, u_i) \in T$ for $1 \leq i \leq n$. Since, for $i \neq j$, $m_i$ and $m_j$ can be the same node in $T$, we can rewrite the above equation as follows:

$$Cost(P, U) = \sum_{m \in T} f'(m, P) \times Cost(m)$$

where $f'(m, P)$ is the number of trajectories anonymized by $P$ using cloak sequence $m$. Since $C$ represents the equivalence class of $P$,

$$\forall m \in T, \quad f'(m, P) = f(m, C)$$

where $f(m, C)$ is as defined in Definition 4. Therefore the cost of configuration $C$ of $T$ can be written as:

$$\begin{aligned} Cost_c(C, U) &= \sum_{m \in T} f(m, C) \times Cost(m) \\ &= \sum_{m \in T} f'(m, P) \times Cost(m) \\ &= Cost(P, U) \end{aligned} \tag{3}$$

□

## A.5 Lemma 5

PROOF. First we give the intuion of this proof. Let $C_u$ be the optimum k-summing configuration for $T_u$. Notice that every node in $T_u$ is also a node in the corresponding $T_{usq}$. Therefore $C_u$ is a valid but not necessarily optimal k-summing configuration for $T_{usq}$. Consequently, the cost of the optimum k-summing configuration for $T_{usq}$ is not more than the cost of $C_u$.

To prove it formally, we can define a configuration $C'$ for $T_{usq}$ as follows

- $C'(m) = C(m)$ for $m \in T_{usq}$ and $m \in T_u$
- $C'(m) = \sum_{l=1}^{4} C'(m_i)$ for $m, m_1 \ldots m_4 \in T_{usq}$ and $m \notin T_u$, where $m_1 \ldots m_4$ are child nodes of $m$

The above conditions ensure that $C'$ only uses those nodes of $T_{usq}$ for anonymization that are also in $T_u$. Each node $m$ that is not in $T_u$ pass-up all the trajectories that are passed up by $m$'s child nodes to be anonymized by $m$'s ancestors. $C'$ is a valid configuration since $\forall m$, $C'(m) < d(m)$ and $C'(m) \leq \sum_{l=1}^{4} C'(m_i)$. $C'$ is k-summing since $C$ is k-summing and $Cost(C')$ = $Cost(C)$ since the newly inserted nodes are not used for anonymization and the nodes that are used for anonymization have the same cost in the two $G$-trees. □

## A.6 Lemma 7

PROOF. Let $U$ be a set of user trajectories, $B$ a USeq-Btree, $C$ an optimal configuration of $B$, and $P$ a policy it represents. Suppose there is a node (cloak sequence of semi-quadrants) $m \in B$ such that $k(h(m)+1) < C(m) < d(m)$. Then by pigeonhold principle, there is a set $S$ of at least $k$ trajectories such that (i) all the trajectories in $S$ are masked by $m$, and (ii) each trajectory in $S$ is anonymized by $P$ using some of the $h(m)$ ancestors of $m$, and (iii) if all the trajectories in $S$ are removed, the cloak sequence they were mapped to under $P$ continue to anonymize at least $k$ trajectories. We then construct a policy $P'$ that anonymize the trajectories in $S$ using $m$ instead of its ancestors. $P'$ continues to be *TP*-aware sender k-anonymous, but has lower cost, contradicting the optimality of $P$. $\square$

## A.7 Theorem 1

PROOF. We prove this by reducing the problem of *optimum k-anonymization of relational tables* on binary alphabet, shown to be NP-hard in [26, 13], to the optimum offline *TP*-aware sender k-anonymity with quad-cloaks.

We first briefly describe the problem of optimum $k$-anonymization of relational tables on binary alphabet with suppression. Let $T$ be a relational table with $m$ columns where each tuple contains data corresponding to a unique user. The tuples of $T$ can be considered to be $m$-dimensional vectors $v_i$ drawn from $\Sigma^m$, where $\Sigma = \{0,1\}$. Thus $T$ can also be represented as a subset $T \subseteq \Sigma^m$. Let $\star$ be a fresh symbol not in $\Sigma$. Suppression is defined as follows:

DEFINITION 6 (**Suppressor**). *Let $f$ be a map from $T$ to $(\Sigma \cup \{\star\})^m$. We say $f$ is a a* suppressor *of $T$ if for all $t \in T$ and $j = 1, \dots, m$ it is the case that $f(t)[j] \in \{t[j], \star\}$.*

Inuitively, a suppressor function replaces the values of certain attributes in certain tuples with $\star$. The idea behind a suppressor function is that by replacing values of certain attributes in a set of tuples with $\star$, it can make all of them identical such that an attacker cannot associate a tuple in that set to the actual user. This is formalized in the following definition.

DEFINITION 7 (**k-Anonymity**). *Let $T$ be a relational table and $f$ a suppressor function. The anonymized table $f(T)$ is said to be $k$-anonymous if for any $t \in T$, there exist $k$ distinct vectors in $T$ such that $f(t_1) = f(t_2) = \cdots = f(t_k) = f(t)$.*

Since there can be many possible suppressor functions, the goal is to find one that minimizes the number of suppressed values i.e. number of $*$ in the anonymized table. The problem of optimum k-anonymization of relational tables on binary alphabets is defined as follows.

DEFINITION 8 (**Optimum k-anonymity**). *Given a relational table $T \subseteq \Sigma^m$, and a positive integer $c \in \mathbb{N}$, is there a suppressor $f$ such that $f(T)$ is k-anonymous, and the total number of vector coordinates suppressed in $f(T)$ is at most $c$?*

For ease of presentation we use k=3 and reduce the optimum 3-anonymity of relational table on binary alphabet to optimum offline *TP*-aware sender 3-anonymity. This particular relational problem we reduce from is proved NP-hard in [26, 13]. Given a relational table $T$ with $n$ $m$-dimensional tuples over binary alphabet $\{0,1\}$, we create a set of $n$ user-history objects with trajectories of length $m$. For each tuple $t_i$, we create a user-history object $u_i$ and set the location at the $j$-th snapshot of the trajectory as

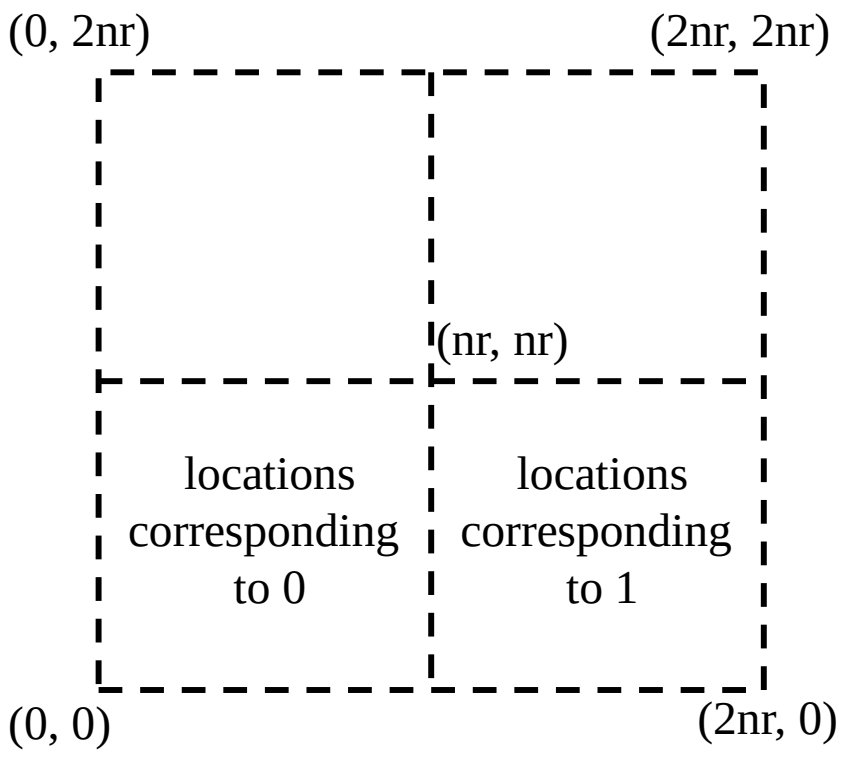

Figure 16: Locations corresponding to the binary data in a Relational Table

- $(i,i)$ if the $j$-th column in $t_i$ has the value 0.
- $(nr+i,i)$ otherwise, where $r = 2^{\lceil \lg \sqrt{2mn} \rceil}$.

We construct a quad-tree $Q$ such that the root node represents the square region $(0,0)$ (left-bottom coordinates) to $(2nr, 2nr)$ (right-top coordinates) as shown in Figure 16. The root quadrant is divided into 4 equal square sub-quadrants. We show that the cost of the optimal 3-ANONYMITY solution for $T$ is at most $c$ if and only if the cost of the optimum policy that provides *TP*-aware sender k-anonymity to the set of users $U$ constructed above is at most $(4c+1)n^2r^2$.

Suppose that there is a solution that finds the optimum quad-cloak policy $P$ of cost at most $(4c+1)n^2r^2$. We construct a suppressor $f$ that k-anonymizes $T$ as follows. For any $1 \le i \le n$ and any $1 \le j \le m$, if the $j$-th location in the trajectory of $u_i$ is masked by the root node of $Q$ in the cloak sequence used to anonymize $u_i$, then $f(t_i)[j] = \star$ and $f(t_i)[j] = t_i[j]$ otherwise. Given the upper bound on the cost of the policy there can be at most $c$ such locations in the trajectories of the users objects that are masked by the root node of $Q$ in the cloak sequences used to anonymize them. Therefore the cost of $f$ is at most $c$. Moreover, since $P$ preserves sender 3-anonymity, there must be 3 trajectories that are anonymized to the same cloak sequence and by construction these 3 trajectories will be anonymized the same way by the suppressor $f$ and hence $f$ is 3-anonymous.

Next let assume $f$ is a suppressor that provides 3-anonymity to $T$ and whose cost is at most $c$. Using $f$ we define a quad-cloak policy $P$ for the set $U$ of user-history objects constructed above. Policy $P$ assigns a quadrant to the $j$-th position of user trajectory $T_i$ by looking up the value of $f(T_i)[j]$:

- If $f(T_i)[j] = \star$, then $P$ uses the biggest quadrant $(0,0)$ to $(nr, nr)$.
- If $f(T_i)[j] = 0$, then $P$ uses a cloak sequence with the quadrant $(0,0)$ to $(n,n)$ at the $j_{th}$ position to anonymize user $u_i$.
- If $f(T_i)[j] = 1$, then $P$ uses a cloak sequence with the quadrant $(nr, 0)$ to $(nr+n, n)$ at the $j_{th}$ position to anonymize user $u_i$.

$P$ is a valid policy as every cloak used masks the corresponding location. For any two tuples $t_a$ and $t_b$, $f(t_a) = f(t_b)$ implies that $P$ uses the same cloak sequence to anonymize users-history objects $u_a$ and $u_b$. Given that $f$ provides 3-anonymity to $T$ there must be 3 users that are anonymized using the same cloak sequence hence $P$ provides *TP*-aware 3-anonymity to the set of users $U$. Furthermore, since the cost of $f$ is at most $c$, there are at most $c$ suppressions and hence at most $c$ locations in the trajectories of the users in $U$ are anonymized by $P$ to the root node of $Q$. The sum of the area of these cloaks is at most $4cn^2r^2$. The remaining locations in the trajectories of users of $U$ are anonymized using cloaks of size $n^2$. Since there are at most $mn$ such locations, the total cost of $P$ is $4cn^2r^2 + mn^3 \leq 4cn^2r^2 + 2mn^3 \leq (4c+1)n^2r^2$. $\square$

## A.8 Theorem 2

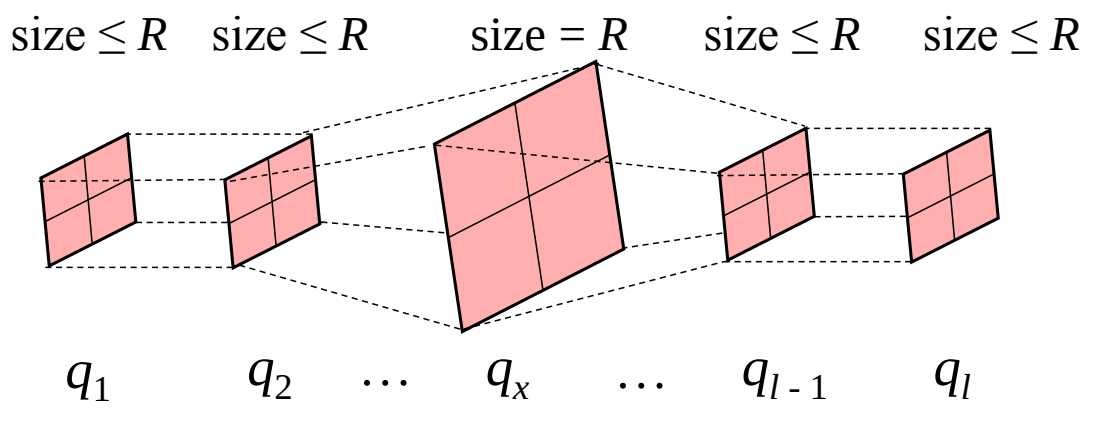

Figure 17: Quad-cloak policy $P$

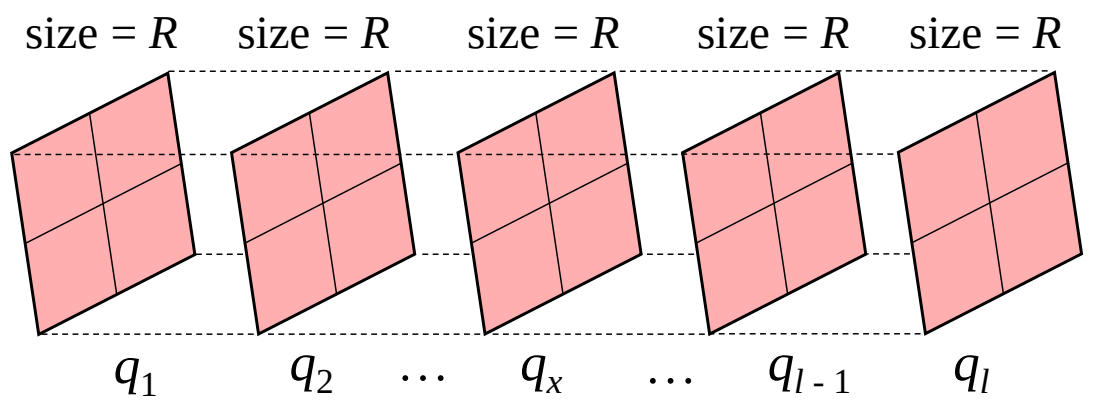

Figure 18: Uniform quad-cloak policy $P'$

PROOF. We prove this theorem by showing that any quad-cloak policy $P$ that provides *TP*-aware sender $k$-anonymity can be transformed to a uniform quad-cloak policy $P'$ that provides the same *TP*-aware sender $k$-anonymity guarantee, and for every input trajectory $u$, its corresponding cloak sequence by $P'(u)$ has a cost that is at most $l$ times the cost of cloak sequence by $P(u)$.

The policy $P'$ is constructed as follows. For each user trajectory $u$, let the cloak sequence of $P(u)$ be $s = \langle q_1, q_2, \cdots, q_l \rangle$ as shown in Figure 17. Let $R$ be the size of the biggest cloak in $s$. Now to construct $P'$, we "expand" each $q_i$ in $s$ to size $R$ as shown in Figure 18. More specifically, let $f_R(x)$ be the lowest ancestor of $x$ or $x$ itself that has size $R$. Then $P'$ will anonymize $u$ using sequence of cloaks $s' = \langle f_R(q_1), f_R(q_2), \cdots, f_R(q_l) \rangle$. In this way, the cost of cloak sequence $s$ is $l \times R$, which is less than $l \times cost(s)$. It then follows that the overall cost of $P'$ is at most $l$ times the overall cost of $P$.

The constructed $P'$ also provides *TP*-aware sender $k$-anonymity since all trajectories that were anonymized to the same cloak sequence by $P$ will now be anonymized to the same cloak sequence by $P'$. Finally, since the above results apply to any quad-cloak policy, they apply to the optimum quad-cloak policy $P_{opt}$ as well. That is, there exists a uniform quad-cloak policy, which is not necessarily the optimum among all uniform quad-cloak policies, that has a cost at most $l$ times the cost of $P_{opt}$. Consequently, the cost of the optimal uniform quad-cloak policy is bounded by $l$ times the cost of $P_{opt}$ as well. $\square$

## A.9 Theorem 3

PROOF. Let $U$ be a set of trajectories. Lemma 3 shows that **Traj-anon** computes optimum uniform quad-tree policy for anonymizing $U$. According to Theorem 2 this optimum solution is $l$-approximation of the optimum quad-tree policy for anonymizing $U$. Therefore, **Traj-anon** computes the $l$-approximation solution to the problem of optimum *TP*-aware sender k-anonymity using quad cloaks. $\square$

## A.10 Theorem 4

PROOF. We prove this by reducing the decision version of optimum policy-aware snapshot k-anonymization with circular cloaks to the decision version optimum offline *TP*-aware k-anonymization with circular cloaks. First we briefly describe the problem of policy-aware snapshot k-anonymization with circular cloaks.

Let $D$ be an instance of location database with schema $S = \{userid, locx, locy\}$ and $SC$ be a set of points in 2-dimensional space. A snapshot policy with circular cloaks is defined as a deterministic function that maps locations in $D$ to circular cloaks, each centered at some point from $SC$, with no restriction on radius. The cost of a snapshot policy with circular cloaks is computed as:

$$Cost_s(P, D) = \sum_{l \in D} Cost(P(D, l))$$

where the cost of the cloak $P(D, l)$ is the area of circular cloak.

DEFINITION 9 (**Snapshot k-anonymity with circular cloaks**). *Given an instance $D$ of location database and $SR$ be the set of points in 2-dimensional space. Is there a snapshot policy $P$ with circular cloaks that provides policy-aware sender k-anonymity and whose $Cost_s(P, D) \leq C$.*

We reduce an instance $I$ of the above problem to an instance $I'$ of the optimum offline *TP*-aware sender k-anonymity with circular cloaks. For each tuple $t \in D$, we create an user-history object $u$ with trajectory of length 1 and set $u.userid() = t.userid$ and $u.location(1) = (t.locx, t.locy)$. Let the resulting set of user-history objects be $U$. We create an instance $I'$ of optimum offline TP-aware k-anonymization with circular cloaks using $U$ and $SR$. We prove that there is snapshot policy $P$ with circular cloaks that provides policy-aware snapshot sender k-anonymity w.r.t. $D$ and $Cost_s(P, D) \leq C$, if and only if there is an anonymization policy $P_t$ that provides $TP$-aware sender k-anonymity solution w.r.t. $U$ and $Cost(P_t, U) \leq C$.

Let $P_t$ be an offline policy that uses circular cloaks and provides *TP*-aware sender k-anonymity to the set $U$ of user-history objects (constructed above). Let the cost of $P_t$ be $Cost(P_t, U) \leq U$. Since the user-objects in $U$ are of length 1, the bundles obtained with $P_t$ are also of length 1. We use $P_t$ to obtain a snapshot policy $P_s$ for $D$ as follows. For each tuple $t \in D$, we define $P_s(D, (t.locx, t.locy)) = b.cloak(1)$ where $b = P_t(U, u)$ for the user-history object $u$ such that $u.userid() = t.userid$. Since $P_t$ provides *TP*-aware sender k-anonymity, there exists at least $k$ user-history objects that are anonymized to a bundle $b$. Therefore the policy $P_s$ as defined above also anonymizes at least $k$ locations to the circular cloak in bundle $b$. Hence $P_s$ provides policy-aware snapshot k-anonymity. Moreover, since $Cost(P_t, U) \leq C$ and $Cost(P_t, U) = Cost_s(P_s, D)$, therefore $Cost_s(P_s, D) \leq C$.

Suppose there exists an snapshot policy $P_s$ with circular cloaks that provides policy-aware sender k-anonymity to $D$ and whose cost $Cost_s(P_s, D) \leq C$. We use $P_s$ to obtain a policy $P_t$ as follows. For every user-history object $u \in U$ corresponding to the tuple $t \in D$ such that $u.userid() = t.userid$, we define $P_t(U, u) = b$ where $b$ is a bundle of length 1 and $b.cloak(1) = P_s(D, (t.locx, t.locy))$. Since $P_s$ provides policy-aware snapshot sender k-anonymity, there exists at least $k$ locations that are aonymized to the same cloak. Therefore there are at least $k$ user-history objects that are anonymized to the same bundle and $P_t$ is *TP*-aware sender k-anonymous. Moreover, since the $Cost_s(P_s, D) \leq C$, and $Cost(P_t, U) = Cost_s(P_s, D)$, therefore $Cost(P_t, U) \leq C$. $\square$